\documentclass[10pt]{article}
\usepackage{amsmath,amssymb,color,epsf}
\usepackage{amssymb,slashed,latexsym,amsmath,multirow,color,dsfont}
\usepackage{graphics}
\usepackage{graphicx}
\usepackage{epstopdf}
\usepackage{cite}

\makeatletter
\@addtoreset{equation}{section}
\makeatother

\usepackage{amsfonts}

\usepackage{float}

\usepackage{soul}

\usepackage{amsfonts}

\usepackage[font=footnotesize,labelsep=newline,labelfont=sc,justification=centering,position=top]{caption}

\newcommand{\bea}{\setlength\arraycolsep{2pt} \begin{eqnarray}}
\newcommand{\eea}{\end{eqnarray}}

\usepackage{hyperref}

\newsavebox{\uuunit}
\sbox{\uuunit}
    {\setlength{\unitlength}{0.825em}
     \begin{picture}(0.6,0.7)
        \thinlines
        \put(0,0){\line(1,0){0.5}}
        \put(0.15,0){\line(0,1){0.7}}
        \put(0.35,0){\line(0,1){0.8}}
       \multiput(0.3,0.8)(-0.04,-0.02){12}{\rule{0.5pt}{0.5pt}}
     \end {picture}}

\usepackage{mathtools}

\DeclarePairedDelimiterX\braket[2]{\langle}{\rangle}{#1 \delimsize\vert #2}

\def\be{\begin{equation}}
\def\ee{\end{equation}}
\def\ba{\begin{array}}
\def\ea{\end{array}}
\def\bea{\begin{eqnarray}}
\def\eea{\end{eqnarray}}
\def\bd{\begin{displaymath}}
\def\ed{\end{displaymath}}

\def\g{\gamma}

\DeclareMathOperator{\Tr}{Tr}

\begin{document}

\begin{titlepage}

\begin{center}

\vskip 1.5cm

{\Large \bf High temperature behavior of non-local observables in  boosted strongly coupled plasma: A holographic study}
\vskip 1cm

{\bf Atanu Bhatta\,$^1$, Shankhadeep Chakrabortty\,$^2$, Suat Dengiz \,$^3$, \\ Ercan Kilicarslan \,$^4$} \\

\vskip 25pt

{\em $^1$ \hskip -.1truecm Centre for High Energy Physics, Indian Institute of Science, ,\\  C.V. Raman Avenue, Bangalore 560012, India.} \\ 

{\em $^2$ \hskip -.1truecm Department of Physics, Indian Institute of Technology Ropar, \\ Rupnagar, Punjab 140001, India.} \\ 

{\em $^3$ \hskip -.1truecm Department of Mechanical Engineering, University of Turkish Aeronautical Association, 06790 Ankara, Turkey. } \\ 

{\em $^4$ \hskip -.1truecm Department of Physics,   Usak University, 64200, Usak, Turkey.  \vskip 5pt }

{email: {\tt atanubhatta@iisc.ac.in, s.chakrabortty@iitrpr.ac.in, sdengiz@thk.edu.tr, ercan.kilicarslan@usak.edu.tr} } \\

\vskip 25pt


\end{center}

\vskip 0.5cm

\begin{abstract}
In this work, we perform a holographic analysis to study non local observables associated to a uniformly \textit{boosted} strongly coupled large $N$  thermal plasma in $d$-dimensions. In order to accomplish the holographic analysis, the appropriate dual bulk theory turns out to be $d+1$ dimensional \textit{boosted} AdS-Schwarzschild blackhole background.  In particular, we compute entanglement entropy of the boosted plasma at high temperature living inside a strip geometry with entangling width $l$ in the boundary at a particular instant of time.  We also study the two-point correlators in the boundary by following geodesic approximation method. For analyzing the effect of boosting on the thermal plasma and correspondingly on both non local observables, we keep the alignment of the width of region of interest both parallel and perpendicular to the direction of the boost.  We find our results significantly modified compared to those in un-boosted plasma up to the quadratic order of the boost velocity $v$. More interestingly, the relative orientation of the boost and the entangling width plays a crucial role to quantify the holographic entanglement entropy in the boundary theory. The breaking of rotational symmetry in the boundary theory due to the boosting of the plasma along a specific flat direction causes this interesting feature.

\end{abstract}
 \end{titlepage}

\tableofcontents


\setcounter{page}{1}
\section{Introduction}
 
 Theoretical understanding of strongly coupled quantum field theories existing in nature, including the recently discovered Quark-Gluon-Plasma (QGP) in relativistic heavy ion collision \cite{ Adcox:2004mh, Adams:2005dq, Back:2004je, Baier:1996kr, Eskola:2004cr, Shuryak:2008eq, Shuryak:2014zxa, Panero, Kozcaz} is hard to achieve by applying the standard technics in perturbation theory. On the other side, $AdS/CFT$ correspondence offers us an indirect way to probe the non-perturbative effects in strongly coupled systems by exploring a suitable dual weakly coupled theory of gravity. A very well-studied example of this correspondence is the duality between type $IIB$ supergravity in $AdS_5 \times S^5$ and strongly coupled large $N$, $\mathcal{N} = 4$ Super Yang-Mills theory living in the four dimensional conformal boundary of the $AdS_5$ \cite{Maldacena:1997re, Gubser, Witten}. 
 Further generalization to this correspondence has been achieved by associating temperature to the boundary gauge theory and by identifying the dual gravity spacetime to be AdS-Schwarzschild black hole \cite{Witten:1998zw}. Considering AdS-Schwarzschild black hole as the dual bulk gravity, the authors in \cite{Fischler:2012ca} have holographically explored the finite temperature behavior of the non-local observables such as entanglement entropy, two-point correlation function and the expectation value of the Wilson loop in the  strongly coupled boundary plasma at finite temperature. Moreover, the closed analytic expressions for those non-local boundary observables have been computed both at high and low temperature regimes of the boundary theory.  

In this work, we consider the boundary theory to be a strongly coupled large $N$ thermal plasma moving with a \textit{uniform boost} $v$ with respect to an observer seating in a static reference frame attached to the flat boundary spacetime (rest frame observer). An example of such strongly coupled theory is $\mathcal{N} = 4$ Super Yang-Mills thermal plasma living in four dimensional flat spacetime.  In the bulk, the gravity background dual to the boosted thermal plasma is realized as a uniformly boosted AdS Schwarzschild planar black hole \cite{Myers_Ent}.\footnote{Note that under a specific long wave width approximation, the quantum dynamics of strongly coupled thermal plasma simplifies to an effective classical dynamics of ideal fluid having well-defined holographic dual described by boosted black brane solution \cite{Bhattacharyya:2008jc}.}  Generally, temperature of the boundary plasma is measured by a boundary observer co-moving with the plasma (co-moving observer). 
However, as our current objective is to explicitly capture the effect of boost parameters, we express the outcome of our analysis in terms of temperature measured by the rest frame observer. Similar set up for obtaining the temperature has been discussed in \cite{  Hubeny:2009rc, Horowitz:1996ay}. Nevertheless, the temperature $T$ as measured by the co-moving observer is related to the temperature  $T_{\text{boost}}$  seen by the rest frame observer in a very simple way:  $T_{\text{boost}} = \frac{T}{\gamma}$. (See appendix-A for derivation). We focus into the high temperature regime of the boosted plasma and estimate the modification of certain nonlocal observables such as entanglement entropy and two-point correlators up to the quadratic order of the boost parameter $v$. In particular, using the uniformly boosted AdS Schwarzschild planar black hole we perform a holographic computation of entanglement entropy of the boosted plasma using a strip geometry with entangling width $l$ in the boundary spacetime. Moreover, we holographically compute the two point correlator of primary operators inserted at two spacetime points separated by a width $l$ lying along one of the flat directions of the boundary theory. The holographic analysis of two point correlator involves the computation of appropriate geodesic width followed by the well-known geodesic approximation method in the bulk spacetime. To emphasize the effect of boost parameter $v$ in carrying out the holographic computation for both non-local observables, we keep the orientation of the width of interest both parallel and perpendicular to the direction of boost. The $v \to 0$ limit of our final results consistently reproduces the findings in \cite{Fischler:2012ca}.  To mention a few similar analysis of non-local observable of a thermal plasma in the high temperature regime we refer \cite{Mozaffara:2016iwm, Karar:2018hvy}.  Similarly, the analytical study of such observables in the low temperature limit is discussed in \cite{Myers_Ent, Fischler:2012ca, Mishra:2016yor, Karar:2019wjb}.
A diametrically opposite limit corresponding to the infinite boost has been also studied \cite{Narayan:2012ks, Narayan:2013qga,  Narayan:2015lka, Mukherjee:2014gia}.

It is important to mention that studying non-local observables is beyond the scope of analytical technics for arbitrary values of boundary parameters ($T_{\text{boost}}, v$). However, such analytical study is perfectly viable within a high temperature limit accompanied by a small boost approximation. We explore the high temperature limit in a systematic way by introducing a dimensionless parameter $T_{\text{boost}} l$, where $T_{\text{boost}}$  and $l$ are the temperature and width of the entangling region as seen by a rest frame observer in the boundary theory. Given the blackhole background and the corresponding Hawking temperature, we can always choose the width of the entangling region very large such that the inequality $T_{\text{boost}} >> 1/l$, signifying the high temperature regime, holds. 
The physical motivation of exploring \textit{only} high temperature regime becomes more evident as one introduces dissipation into the boundary theory. Usually, a dissipative system attains a local thermodynamical equilibrium where the temperature is a slowly varying function of spacetime and the inverse of the temperature sets a local length scale in the theory. In such a situation, one needs to take the width of the entangling region to be very large as compared to the local length scale so that the dissipative characteristics can be captured. Note that,  in \cite {Myers_Ent, Fischler:2012ca}, by virtue of equivalence between the temperature scale and the length scale, for ideal fluid at finite temperature, the analytical expression of entanglement entropy has been achived by considering the entangling region small. 

The lay-out of paper is as follows: In section (\ref{s1}) we discuss the holographic computation of the entanglement entropy and its high temperature limit. Here, we elaborately discuss the correction coming due to the uniform boost applied to the thermal plasma. We make two separate analysis for parallel and perpendicular cases to emphasize the effect of introducing boost in the thermal plasma.  In section (\ref{s2}), we study the holographic analysis of two point correlator and also its high temperature behavior. We also give systematic derivation of the modification of the correlators due to the uniform boost. Finally,  in section (\ref{s3}), we conclude by mentioning our results and discuss some future directions. 
 

\section{Entanglement entropy in a boosted plasma}\label{s1}

The idea of quantum entanglement indicates that a quantum mechanical measurement on a component of an entangled pair can indeed affect the outcome of a measurement on the other component of the pair. The correlation between the entangled pair is inherently nonlocal and unlike the classical correlation, it depends on the measurement itself. A well-defined measure of quantum entanglement can be used as a suitable non-local probe to explore various interesting phases of a physical system. In the present analysis, among various measures of entanglement, we consider entanglement entropy (EE) as a suitable non-local probe to explore the strong coupling phase of a large $N$ thermal plasma. 

To define EE quantitatively, let us proceed with a quantum mechanical bipartite system for which the Hilbert space is defined as 
\begin{equation}
\mathcal{H} = \mathcal{H}_A \otimes \mathcal{H}_B,
\end{equation}
where $ \mathcal{H}_A $ and $ \mathcal{H}_B$ are the Hilbert space of the individual subsystems $A$ and $B$. To evaluate the EE first we need to construct reduced density matrix by taking partial trace of total density matrix $\rho$ over $\mathcal{H}_B$ as
\begin{eqnarray}
\rho_{A} = \Tr_{B}\, \rho.
\end{eqnarray}
Then, the von-Neumann entropy associated to the reduced density matrix becomes the EE of the system $A$ 
\begin{eqnarray}\label{eq:EElog}
S_A = - \Tr_{A}\,\Big( \rho_{A}\, \log \rho_{A}\Big).
\end{eqnarray}

It is very difficult to implement the aforementioned prescription (\ref{eq:EElog}) to compute EE in perturbative quantum field theory (QFT) in arbitrary dimensions. In two dimensional QFT preserving conformal invariance, the replica trick method turns out to be very useful in obtaining the results for EE \cite{  Holzhey:1994we, Calabrese:2004eu}. The EE in this case contains  finite non-local contributions as well as local divergent part which is regularized by an appropriate UV cut-off \cite{Calabrese:2004eu, Casini:2009sr}.

In strongly coupled QFT, it is still not well-understood how to compute the EE directly by using (\ref{eq:EElog}). Ryu and Takayanagi (RT) conjectured a new holographic prescription which associates the EE of the boundary field theory endowed by a conformal symmetry structure, with the area of an extremum hyper-surface, a purely geometrical quantity in the dual bulk gravity \cite{Ryu:2006bv,Ryu:2006ef,Nishioka:2009un}. In particular, entanglement entropy of a region~$A$ in $d$-dimensional strongly coupled boundary theory is conjectured to be 
\begin{equation}\label{EE}
S_A = \frac{{\rm Area} (\g_A)}{4 G_N^{(d+1)}},
\end{equation}
where $\gamma_A$ is a co-dimension two space-like minimal surface  in the holographically dual $d+1$ dimensional bulk gravity,  $G_N^{(d+1)}$ signifies the Newton constant in the $d+1$ bulk spacetime. The boundary of the co-dimension two space-like minimal surface, ${\partial \gamma}_A$ coincides with the boundary $\partial A$ of the entangling region $A$. The aforementioned equation (\ref{EE}) serves as our working formula to compute entanglement entropy of a sub-region $A$ in a strongly coupled boosted thermal plasma.

The $d+1$ dimensional gravity background dual to large $N$, strongly coupled  plasma at finite temperature  living in $d$ dimensional boundary is given as the following AdS Schwarzschild black hole spacetime, 
\begin{equation}
 ds^2 = \frac{r^2}{R^2}\left[ -(1-\frac{r_H^d}{r^d})dt^2 + 
 dx^2 + d\vec{x}^2_{d-2}+\frac{R^4}{r^4}\frac{dr^2}{1-\frac{r_H^d}{r^d}}\right].
 \label{unboostedmetric}
\end{equation}
In the present analysis, we introduce a uniform boost $v$ to the thermal plasma along a spatial flat direction in the boundary, say x. By virtue of $AdS/CFT$ duality, the holographic dual of the uniformly boosted plasma can be described as boosted AdS Schwarzschild black hole spacetime,
\begin{equation}
 ds^2 = \frac{r^2}{R^2}\left[ -dt^2 + 
 dx^2 +\gamma^2 \frac{r_H^d}{r^d}\left( dt+v dx\right)^2+ d\vec{x}^2_{d-2}+\frac{R^4}{r^4}\frac{dr^2}{1-\frac{r_H^d}{r^d}}\right], 
 \label{boostedmetric}
\end{equation}
with $\gamma = 1/\sqrt{1-v^2}$.
In natural unit, the boost velocity $v$ is a dimensionless parameter and its value is bounded within [0,1]. 
\subsection{Holographic computation in parallel case}
To understand the effect of boost on the entanglement structure of the thermal plasma we  first consider the entangling region $A$ to be a strip in the boundary defined in a constant time slice $(t=t_0)$ as, 											

\begin{equation}
 x\in\left[-\frac{l}{2},\frac{l}{2}\right]; \quad x^i \in \left[-\frac{L}{2},\frac{L}{2}\right] \quad (i=1,2,\dots d-2),
\end{equation}
where $x$ and $x^i$s are the spatial coordinates in the boundary theory.  We also take $L\rightarrow \infty$ so that the entangling strip region looks symmetrical with respect to all $x^i$ directions. Note that in the present set up the direction of boost velocity of thermal plasma and the alignment of the entangling width $l$ are both along $x$ direction and we call it as $parallel$ case.

A suitable ansatz for a co-dimension two space-like surface $\gamma_A$ embedded in the $d+1$ dimensional dual gravity theory can be parameterized by $d-2$ number of coordinates $\sigma^{\alpha}, \alpha = 1,2,.....,d-2$. The choice of those coordinates is $\sigma^1= x, \sigma^{i} = x^i, i = 1,2,.....,d-3$.
As we impose the limit $L\rightarrow \infty$, among all the coordinates on $\gamma_A$ the only non trivial profile can be assigned to $x = x(r)$.
 The induced metric on surface $\gamma_A$  reads as,
\begin{align}
 G_{xx} &= \frac{r^2}{R^2} \Big[ \left(1+ \frac{r_H^d}{r^d}\gamma^2 v^2\right)+ \frac{r'^2 R^4}{r^4}\left(1-\frac{r_H^d}{r^d}\right)^{-1}\Big] \nonumber\\ 
 G_{ii} &= g_{ii} = r^2 \quad \forall i \in (1,2,\dots d-2),
\end{align}
with $r' \equiv \frac{dr}{dx}$.  
The corresponding area functional we aim to minimize takes the form, 
\begin{align}
\label{eefunctional}
 \mathcal{A}^{||} = \frac{L^{d-2}}{R^{d-2}} \int dr r^{d-2} {\left[ \frac{1}{R^2}\left(r^2 + \gamma^2 v^2 \frac{r_H^d}{r^{d-2}} \right) x'^2 
 + \frac{R^2}{r^2} \left(1-\frac{r_H^d}{r^d}\right)^{-1}\right]}^{\frac{1}{2}},
\end{align}
with $x' \equiv \frac{dx}{dr}$. The fact that the action (\ref{eefunctional}) which has an explicit dependence of boost parameter, turns out to be a key feature from the perspective of our present analysis. It is also important to mention that $v \to 0$ limit can be smoothly taken in (\ref{eefunctional}) and also in all subsequent steps of our computation to reproduce the known results for un-boosted case \cite{Fischler:2012ca}. 

 The minimization procedure boils down to the computation of the on-shell area functional. Since the area functional does not have any explicit dependence on the variable function $x(r)$, we can detour the process of obtaining equation of motion by constructing a much simpler object, the first integral of motion
\begin{equation}
  \frac{ r^{d-2}\Big[\frac{1}{R^2}\left(r^2+\gamma^2 v^2 \frac{r_H^d}{r^{d-2}}\right) x' \Big]}{\sqrt{ \frac{1}{R^2} \left(r^2+\gamma^2 v^2 \frac{r_H^d}{r^{d-2}}\right)x'^2 + \frac{R^2}{r^2}\left(1-\frac{r_H^d}{r^d}\right)^{-1}}} = C,
\end{equation}
where $C$ is some arbitrary constant needed to be fixed by imposing suitable boundary condition. A natural choice of boundary condition fitting with the geometry of the surface $\gamma_A$ is 
\begin{equation}
\lim_{x' \to \infty}  r=r_t^{||},
 \label{eebc}
\end{equation}
where $r_t^{||}$ signifies the radial value of the turning point as $\gamma_A$ approaches deep in to the bulk space time. Using the above mentioned boundary condition (\ref{eebc}) we can re-express the arbitrary constant $C$ in terms of $r_t^{||}$, 
\begin{equation}
 C= {r_t^{||}}^{d-2}\sqrt{\frac{1}{R^2}\Big({r_t^{||}}^2+\gamma^2 v^2 \frac{r_H^d}{{r_t^{||}}^{d-2}}\Big)}.
\end{equation}

With the use of first integral of motion the equation of motion obtained by extremizing the classical action (\ref{eefunctional}) turns out to be a first order differential equation which reads as,
\begin{equation}
 \frac{dx}{dr}= \pm \frac{R^2  {r_t^{||}}^{d-2}\sqrt{{r_t^{||}}^2+\gamma^2v^2\frac{r_H^d}{{r_t^{||}}^{d-2}}}\left(1-\frac{r_H^d}{r^d}\right)^{-1/2}}{r^{d-1}\left(r^2+\gamma^2v^2\frac{r_H^d}{r^{d-2}}\right)\sqrt{1-\frac{{r_t^{||}}^{2(d-2)}}{r^{2(d-2)}}\frac{{r_t^{||}}^2+\gamma^2v^2\frac{r_H^d}{{r_t^{||}}^{d-2}}}{r^2+\gamma^2v^2\frac{r_H^d}{r^{d-2}}}}}.
 \label{eeeom}
\end{equation}
The co-dimension two hyper-surface approaching inside the bulk has two independent branches and those two branches are smoothly joined at $r=r_t^{||}$.  The corresponding boundary conditions satisfied by these two independent branches are
\begin{eqnarray}
\lim_{r \to \infty} x(r) = \pm \frac{l^{||}}{2}.
\label{neweebc}
\end{eqnarray}
Using any one of the above boundary conditions, we solve the equation of motion (\ref{eeeom}) to relate the free bulk parameter $r_t$ with the free parameter $l$ in the boundary theory
\begin{eqnarray}
 \frac{l^{||}}{2} 
 &=& R^2 \int_{r_t^{||}}^{\infty} dr \frac{{r_t^{||}}^{d-1}\sqrt{1+\gamma^2v^2 \alpha^d}\left(1- \alpha^d\frac{{r_t^{||}}^d}{r^d}\right)^{-1/2}}{r^{d+1}\left(1+\gamma^2v^2\alpha^d\frac{{r_t^{||}}^d}{r^d}\right)\sqrt{1-\frac{{r_t^{||}}^{2(d-1)}}{r^{2(d-1)}}\frac{1+\gamma^2v^2\alpha^d}{1+\gamma^2v^2 \alpha^d\frac{r_H^d}{r^d}}}},
 \label{eesol}
\end{eqnarray}
where  $\alpha  = \frac{r_H}{r_t^{||}}$ is a dimensionless parameter which takes value within (0,1). With a suitable change of variable $u = \frac{r_t^{||}}{r} $, the integral takes the following form,
\begin{equation}
 \frac{l^{||}}{2}=\frac{R^2}{r_t^{||}} \int_0^1 du\sqrt{1+\gamma^2 v^2 \alpha^d} ~u^{d-1} \frac{\left(1-\alpha^d u^d\right)^{-1/2}} {\left(1+\gamma^2v^2\alpha^d u^d\right)\sqrt{1-u^{2(d-1)}\frac{1+\gamma^2v^2\alpha^d}{1+\gamma^2v^2\alpha^d u^d}}}.
 \label{eeu}
\end{equation}

The exact evaluation of the above integral ($\ref{eesol}$) for arbitrary non zero values of $v$ is hard to achieve. Hence, we expand the integrand as a power series of  $v$ and evaluate the integral order by order in the power of $v$. To ensure the convergence of the power series in $v$, we always assume $v<1$. Note that in natural unit, choosing $v < 1$ actually implies a demarcation from the full relativistic consideration in the boundary theory. Even then, we can consider $v < 1$ as a first approximation and incorporate the modification coming purely from the boost. A peculiarity in the boost expansion of the integrand shows that all non-vanishing contributions come with the even power of $v$. Here, for our present analysis, we restrict the the boost expansion up to quadratic order in $v$  
\begin{eqnarray}
&&  \frac{l^{||}}{2} = \frac{R^2}{r_t^{||}} \int_0^1 du  \Big[ \frac{1}{(1-u^{2(d-1)})^{1/2}} - v^2 \alpha^d \frac{(2u^d -u^{3d-2} -1)}{2{(1-u^{2(d-1)})}^{3/2} } \Big]u^{(d-1)} \left(1-\alpha^d u^d\right)^{-1/2}. \nonumber \\
 \label{eqn1}
 \end{eqnarray}

Further, we always assume that the two independent branches of co-dimension two hyper-surface always grow inside the bulk in such a way that $r_t$ is always greater than the $ r_H $ ($\alpha<1$). Under this consideration, the $\left(1-\alpha^d u^d\right)^{-1/2}$ factor in the integrand as given in (\ref{eqn1}) is further expanded into a power series in $\alpha$. Finally, we integrate the right hand side of the (\ref{eqn1}) order by order up to quadratic power of $v$. By focusing on the first leading term we obtain \footnote{We use $\frac{1}{\sqrt{1-\alpha^d u^d}} = \sum_{n=0}^{\infty} \frac{\Gamma(n+\frac{1}{2})}{\Gamma(n+1)\Gamma(\frac{1}{2})}\alpha^{nd}u^{nd}$  and 
$ \int_0^1 dx x^{\mu-1} {(1-x^\lambda)}^{\nu-1} = \frac{B (\frac{\mu}{\lambda},\nu) }{\lambda}  = \frac{\Gamma(\frac{\mu}{\lambda}) \Gamma(\nu)}{\lambda ~ \Gamma(\frac{\mu}{\lambda} + \nu)} .$},

\begin{eqnarray}
 {\Big(\frac{l^{||}}{2}\Big)}_{\mathcal{O}(1)} &&= \frac{R^2}{r_t^{||}}  \sum_{n=0}^{\infty} \frac{\Gamma(n+\frac{1}{2})}{\Gamma(n+1)\Gamma(\frac{1}{2})}\alpha^{nd} \int_0^1 du \left[ \frac{u^{(d(n+1)-1)} }{(1-u^{{}2(d-1)})^{1/2}}\right] \nonumber \\
&& =\frac{R^2}{r_t^{||}}  \sum_{n=0}^{\infty} \frac{\alpha^{nd}}{nd+1} \frac{\Gamma(n+\frac{1}{2})\Gamma(\frac{d(n+1)}{2(d-1)})}{\Gamma(n+1)\Gamma(\frac{(nd+1)}{2(d-1)})}.
  \label{eqn2}
 \end{eqnarray}
Note that the above expression (\ref{eqn2}) reproduces the relation between $l$ and $r_t^{||}$ in the unboosted case \cite{Fischler:2012ca}.

The next non-vanishing leading order term is quadratic in the boost velocity $v$,   
\begin{eqnarray}
& {\Big(\frac{l^{||}}{2}\Big)}_{\mathcal{O}(v^2)}  =   \frac{ v^2 \alpha^d R^2}{r_t^{||}}  \sum_{n=0}^{\infty} \alpha^{nd}  \frac{\Gamma(n+\frac{1}{2})}{\Gamma(n+1)} \Big\{  
 \frac{2 \Gamma[\frac{d(n+2)}{2(d-1)}]}{(d(n-1)+3)\Gamma[\frac{d(n-1)+3}{2(d-1)}]}   -  \frac{ \Gamma[\frac{d(n+4)-2}{2(d-1)}]}{{(d(n+1)+1)\Gamma[\frac{d(n+1)+1}{2(d-1)}]}} \nonumber \\
&  -  \frac{ \Gamma[\frac{d(n+1)}{2(d-1)}]}{{(d(n-2)+3)\Gamma[\frac{d(n-2)+3}{2(d-1)}]}} \Big\}.
\label{eqn3}
 \end{eqnarray}

For sufficiently large $n$, all terms inside the curly bracket in (\ref{eqn3}) conspire with each other in such a way that they do not contribute to any new divergence. Now, by collecting first few leading order terms we get a form of $\frac{l}{2}$,
\begin{eqnarray}
\nonumber
&&  \frac{l^{||}}{2} =\frac{R^2}{r_t^{||}}  \sum_{n=0}^{\infty} \frac{\alpha^{nd}}{nd+1} \frac{\Gamma(n+\frac{1}{2})\Gamma(\frac{d(n+1)}{2(d-1)})}{\Gamma(n+1)\Gamma(\frac{(nd+1)}{2(d-1)})}\\ \nonumber
&&  +\frac{R^2  v^2 \alpha^d}{r_t^{||}}  \sum_{n=0}^{\infty} \alpha^{nd}  \frac{\Gamma(n+\frac{1}{2})}{\Gamma(n+1)} \Big[  
 \frac{2 \Gamma[\frac{d(n+2)}{2(d-1)}]}{(d(n-1)+3)\Gamma[\frac{d(n-1)+3}{2(d-1)}]}  -  \frac{ \Gamma[\frac{d(n+4)-2}{2(d-1)}]}{{(d(n+1)+1)\Gamma[\frac{d(n+1)+1}{2(d-1)}]}}  -  \\ 
 && \frac{ \Gamma[\frac{d(n+1)}{2(d-1)}]}{{(d(n-2)+3)\Gamma[\frac{d(n-2)+3}{2(d-1)}]}} \Big] + \mathcal{O}(v^4).
 \label{eqngeneral1}
 \end{eqnarray}

As we substitute $x'$ as given in the equation (\ref{eeeom}) into (\ref{eefunctional}), we get the action functional corresponding to the both branches of co-dimension two minimal surface $\gamma_A$,
\begin{eqnarray}
\label{areaee}
 \mathcal{A}^{||} &=& 2 \frac{L^{d-2}}{R^{d-3}} \int_{r_t^{||}}^{\infty} dr~ r^{d-3} \frac{1}{\sqrt{(1-\frac{{r_t^{||}}^{2(d-1)}}{r^{2(d-1)}}\frac{1+\gamma^2v^2 \frac{r_H^d}{{r_t^{||}}^d}}{1+\gamma^2v^2\frac{r_H^d}{r^d}})}}\left(1-\frac{r_H^d}{r^d}\right)^{-1/2}. \nonumber \\
\end{eqnarray}
 It is evident from the explicit form of integrand in (\ref{areaee}) that the area has a divergence as $r \to \infty$.  To regulate the divergence, we introduce an IR cut-off $r_0$ in the divergent piece of the area functional in the bulk theory.
\begin{eqnarray}
{ \mathcal{A} }^{||}_{\text{infinite}} = \frac{2}{(d-2)} \frac{L^{d-2}}{R^{d-3} } r_0^{d-2}.
\label{divarea}
\end{eqnarray}
By the virtue of holographic duality the IR cut-off $r_0$ in the bulk corresponds an UV cut-off $\delta^{||} = \frac{R^2}{r_0}$ in the boundary theory.
\begin{eqnarray}
{ \mathcal{A} }^{||}_{\text{infinite}}= \frac{2}{d-2} \frac{L^{d-2} R^{d-1} }{{\delta^{||}}^{d-2}}.
\label{divarea1}
\end{eqnarray}
Note that the strongest contribution of entanglement between a region $A$ and region $B$ comes from the boundary $\partial A$ \cite{Bombelli:1986rw, Srednicki:1993im, Casini:2003ix, Das:2005ah}. Consistent to this fact, also in the holographic computation, we see that the UV divergent term turns out to be proportional to the dimension of area in the boundary spacetime. 
Finally we re-expressing the finite part of the area functional in terms of a suitable dimensionless variable $u = \frac{r_t}{r}$,
as follows,
\begin{eqnarray}
\label{finalfinite}
&& { \mathcal{A} }^{||}_{\text{finite}} =   2 {r_t^{||}}^{d-2}\frac{L^{d-2}}{R^{d-3}} \int_{0}^{1} du~ \Big[ \frac{1}{u^{d-1}\sqrt{1-u^{2(d-1)}} }  
+  v^2 \alpha^d  \frac{u^{d-1} (1-u^d)}{2{(1-u^{2(d-1)})}^{3/2}}\Big]\left(1-\alpha^du^d\right)^{-1/2} - \nonumber \\ 
&&  \frac{2}{(d-2)} \frac{L^{d-2}}{R^{d-3} } r_0^{d-2}\nonumber \\
&& = 2 \frac{L^{d-2}}{R^{d-3}} {r_t^{||}}^{d-2} \Big[ \Big \{ \frac{\sqrt{\pi} \Gamma(-\frac{d-2}{2(d-1)})}{2(d-1) \Gamma(\frac{1}{2(d-1)})}  + \sum_{n=1}^{\infty}   \frac{\Gamma(n+\frac{1}{2}) \alpha^{nd}}{\Gamma(n+1)} \frac{1}{2(d-1)} \frac{\Gamma(\frac{d(n-1)+2}{2(d-1)})}{\Gamma(\frac{nd+1}{2(d-1)})} \Big\}    \nonumber \\    
&&  - v^2 \alpha^d  \Big\{ \sqrt{\pi} \Big( \frac{ \Gamma(\frac{d}{2(d-1)})}{2(d-1) \Gamma(\frac{1}{2(d-1)})}  - \frac{ \Gamma(\frac{d}{(d-1)})}{2(d-1) \Gamma(\frac{d+1}{2(d-1)})}\Big) \nonumber \\
&& + \sum_{n=1}^{\infty}   \frac{\Gamma(n+\frac{1}{2}) \alpha^{nd}}{\Gamma(n+1)} \frac{1}{2(d-1)}  \Big(   \frac{\Gamma(\frac{d(n+1)}{2(d-1)})}{\Gamma(\frac{nd+1}{2(d-1)})} - \frac{\Gamma(\frac{d(n+2)}{2(d-1)})}{\Gamma(\frac{(n+1)d+1}{2(d-1)})}\Big) + \mathcal{O}(v^4, \epsilon^2) \Big\} \Big]
 \end{eqnarray}

The expression of ${\mathcal{A} }_{\text{finite}} $ contains an infinite series which is convergent within the regime of the inequality $ r_t^{||} > r_H$. Once the finite part of the area is obtained, by following the RT proposal (\ref{EE}), one can finally calculate the entanglement entropy of the strip in the boundary theory. To express the entanglement entropy in terms of boundary parameters, one needs to solve (\ref{eqngeneral1}) for $r_t^{||}$. Achieving such solution by analytical method for arbitrary values of temperature and boost is not possible.  However, in high temperature region, the analytic expression of the  $r_t^{||}$ turns out to be obtainable order by order in the power of boost parameter $v$.
It is important to note that as the rest frame observer sees the boundary plasma moving with a uniform boost, the realization of entanglement to him/her is based on instantaneous observation. At each instant, the rest frame observer expects some instantaneous correlation between a part of the plasma momentarily confined inside region $A$ in the boundary and the rest of the plasma outside that region. There is no a priori reason to assume that the strength of the entanglement as seen by the rest frame observer varies from one instant to another. Similarly, in the bulk, space-like hyper-surface we consider is defined by some constant time slice in the bulk metric and the corresponding holographic EE for strongly coupled boosted plasma is being computed at that particular instant of time. Since the result should be independent of the choice of constant time slice, we expect the same value for EE for all instants of time.   

In the following, we compute the high temperature behavior $(l^{||} T_{\text{boost}} >> 1)$ of the EE in the boundary theory.  As elucidated previously, within $(l^{||}T_{\text{boost}} >> 1)$ limit,  the extremal surface approaches to cover a part of the horizon ($r_t^{||} \to r_H$). To capture the high temperature limit we re-write the area functional in the following way such that we can avoid divergence in the computation

 \begin{eqnarray}
&&   \mathcal{A}^{||}_{\text{finite}} = 2 \frac{L^{d-2}}{R^{d-3}}  {r_t^{||}}^{d-2} \Big [ \frac{l^{||} r_t^{||}}{2R^2} -  \frac{(d-1)\sqrt{\pi}\Gamma[\frac{d}{2(d-1)}]}{(d-2) \Gamma[\frac{1}{2(d-1)}]}    \sum_{n=1}^{\infty} \frac{\Gamma(n+\frac{1}{2}) \alpha^{nd}}{\Gamma(n+1)} \frac{1}{nd+1} \Big( \frac{d-1}{d(n-1)+2} \Big) \frac{\Gamma[\frac{d(n+1)}{2(d-1)}]}{\Gamma[\frac{nd+1}{2(d-1)}] } \nonumber \\
&& - v^2 \alpha^d \sum_{n=0}^{\infty} \alpha^{nd}\frac{\Gamma(n+\frac{1}{2})}{ \Gamma(n+1)} \Big[\frac{\Gamma(\frac{d(n+1)}{2(d-1)})}{2(d-1)\Gamma(\frac{nd+1}{2(d-1)})} - \frac{\Gamma(\frac{d(n+2)}{2(d-1)})}{2(d-1)\Gamma(\frac{(n+1)d+1}{2(d-1)})} + \frac{2 \Gamma[\frac{d(n+2)}{2(d-1)}]}{(d(n-1)+3)\Gamma[\frac{d(n-1)+3}{2(d-1)}]}  \nonumber \\
&&  -  \frac{ \Gamma[\frac{d(n+4)-2}{2(d-1)}]}{{(d(n+1)+1)\Gamma[\frac{d(n+1)+1}{2(d-1)}]}}  -  \frac{ \Gamma[\frac{d(n+1)}{2(d-1)}]}{{(d(n-2)+3)\Gamma[\frac{d(n-2)+3}{2(d-1)}]}} \Big ] + \mathcal{O}(v^4).
 \label{approxA2rw1}
 \end{eqnarray}
                    
Note that  infinite series present in (\ref{approxA2rw1}) does not give rise to any new divergent term. Now by imposing the limit $r_t^{||} \to r_H$ in (\ref{approxA2rw1}), one can obtain the leading behavior of the minimal area as, 
\begin{eqnarray}
  \mathcal{A}^{||}_{\text{finite}} & = \frac{V_{d-1} r_H^{d-1}}{R^{d-1}} \Big[  1 + \frac{2 R^2}{l^{||} r_H} (\mathcal{S}^{||}_0 + v^2  \mathcal{S}^{||}_1)\Big] + \mathcal{O}(v^4),
  \label{noepsfinite}
  \end{eqnarray}
  where we have denoted the spatial volume of the rectangular strip as $V_{d-1} = l^{||}L^{d-2}$ and $\mathcal{S}^{||}_0$ and $\mathcal{S}^{||}_1$ can be expressed as,
  \begin{eqnarray}
\mathcal{S}^{||}_0  &=& \Big(  -  \frac{(d-1)\sqrt{\pi}\Gamma[\frac{d}{2(d-1)}]}{(d-2) \Gamma[\frac{1}{2(d-1)}]}  +  \sum_{n=1}^{\infty} \frac{\Gamma(n+\frac{1}{2}) }{\Gamma(n+1)} \frac{1}{nd+1} \Big( \frac{d-1}{d(n-1)+2} \Big) \frac{\Gamma[\frac{d(n+1)}{2(d-1)}]}{\Gamma[\frac{nd+1}{2(d-1)}] }  \Big), \nonumber \\
\mathcal{S}^{||}_1 &=&   -\sum_{n=0}^{\infty} \frac{\Gamma(n+\frac{1}{2})}{ \Gamma(n+1)} \Big[\frac{\Gamma(\frac{d(n+1)}{2(d-1)})}{2(d-1)\Gamma(\frac{nd+1}{2(d-1)})} - \frac{\Gamma(\frac{d(n+2)}{2(d-1)})}{2(d-1)\Gamma(\frac{(n+1)d+1}{2(d-1)})} + \frac{2 \Gamma[\frac{d(n+2)}{2(d-1)}]}{(d(n-1)+3)\Gamma[\frac{d(n-1)+3}{2(d-1)}]}  \nonumber \\
&&  -  \frac{ \Gamma[\frac{d(n+4)-2}{2(d-1)}]}{{(d(n+1)+1)\Gamma[\frac{d(n+1)+1}{2(d-1)}]}}  -  \frac{ \Gamma[\frac{d(n+1)}{2(d-1)}]}{{(d(n-2)+3)\Gamma[\frac{d(n-2)+3}{2(d-1)}]}} \Big ] .
\label{paracoeff}
  \end{eqnarray}
Finally following the holographic prescription (\ref{EE}), within the high temperature limit, EE of the uniformly boosted thermal plasma in terms of the temperature $T_{\text{boost}}$ from the point of view of the rest frame observer can be expressed as follows,

   \begin{eqnarray}
  S^{||} = \frac{R^{d-1}}{4 G_N^{d+1}} \Big [ \frac{2}{d-2} {\Big(\frac{L}{\delta^{||}}\Big)}^{d-2} +
 V_{d-1} {\Big(\frac{4 \pi T_{\text{boost}}}{d}\Big)}^{d-1} \{ 1+ v^2\Big(\frac{d-1}{2}\Big)\}  \nonumber \\
 + A_{d-2} {\Big(\frac{4 \pi T_{\text{boost}}}{d}\Big)}^{d-2} \{ \mathcal{S}^{||}_0 + v^2 \Big(\mathcal{S}^{||}_1  + \frac{d-2}{2} \mathcal{S}^{||}_0\Big) \} \Big] + \mathcal{O}(v^4), 
 \label{entropyfinal1}
  \end{eqnarray}

where $A_{d-2} = 2 L^{d-2}$ is the spatial area of the rectangular strip.
Note that the leading contributions of the finite part of EE is proportional to volume whereas the sub-leading part is proportional to the area. In both the finite terms there are respective modifications arising from the boost parameter.

  It is curious to learn about the holographic $c$-function to estimate the possible modification of degrees of freedom due to the emergence of boost parameter in the dual thermal plasma. Following \cite{Myers:2012ed}, we compute the  holographic $c$-function of the boosted plasma, 

  \begin{equation}
  {\mathcal{C}}^{||}_{v} = \mathcal{C}_{v=0} \gamma^{d-1},
  \end{equation}
where $\mathcal{C}_{v=0} $ is the $c$-function for un-boosted plasma.  It is evident from the above expression that the boost enhances the degrees of freedom as compared to that in the un-boosted case.

 So far we are using the approximation $r_t^{||} \to r_H$ in the computation of holographic entanglement entropy for a boosted plasma and it indicates that however close the extremal surface with respect to the horizon is, it can not exactly reach up to the horizon. In the following analysis, we shall assume $r_t^{||} = (1+ \epsilon) r_H$ with $\epsilon << 1$, so that $r_t^{||} \to  r_H$ can be interpreted as a first approximation and the $\epsilon$ parameter measures how fast the extremal surface approaches to the horizon. Up to linear order  $\mathcal{O}(\epsilon)$, we can write, $\alpha  = \frac{1}{1+ \epsilon} \approx 1 - \epsilon$ and rewrite the expression for the entangling width $l^{||}$ (\ref{eqngeneral1}) as follows,
 \begin{eqnarray}
  \frac{l^{||}}{2} &=& \frac{R^2}{(1+ \epsilon) r_H} \frac{\sqrt{\pi}\,\Gamma\left( \frac{1}{2(d-1)}\right)}{\Gamma\left( \frac{1}{2(d-1)}\right)} -\frac{R^2}{(1+ \epsilon) r_H} \frac{1}{\sqrt{2d(d-1)}}\log\left(1-\frac{1}{(1+\epsilon)^d}\right) \cr 
  &+&\frac{R^2}{(1+ \epsilon) r_H} \sum_{n=1}^{\infty}\left[ \frac{1}{nd+1} \frac{\Gamma(n+\frac{1}{2})\Gamma(\frac{d(n+1)}{2(d-1)})}{\Gamma(n+1)\Gamma(\frac{(nd+1)}{2(d-1)})}-\frac{1}{n\sqrt{2d(d-1)}} \right]\frac{1}{(1+\epsilon)^{nd}} \cr 
 &+&\!\!\! \frac{R^2  v^2 (1+\epsilon)^{-d}}{(1+ \epsilon) r_H}  \sum_{n=0}^{\infty}  \frac{\Gamma(n+\frac{1}{2})}{\Gamma(n+1)} \left[  
 \frac{2 \Gamma[\frac{d(n+2)}{2(d-1)}]}{(d(n-1)+3)\Gamma[\frac{d(n-1)+3}{2(d-1)}]}  -  \frac{ \Gamma[\frac{d(n+4)-2}{2(d-1)}]}{{(d(n+1)+1)\Gamma[\frac{d(n+1)+1}{2(d-1)}]}} \right.    \cr  
 && \qquad \qquad \qquad\qquad  - \left. \frac{ \Gamma[\frac{d(n+1)}{2(d-1)}]}{{(d(n-2)+3)\Gamma[\frac{d(n-2)+3}{2(d-1)}]}} \right] \frac{1}{(1+\epsilon)^{nd}} + \mathcal{O}(v^4).
 \label{eqngeneral2}
 \end{eqnarray}
Rearranging both sides of the above expression we get,
\begin{eqnarray}
\frac{1}{\sqrt{2d(d-1)} }\log (d \epsilon) &=& -\frac{l^{||} r_H}{2 R^2} +  \frac{\sqrt{\pi} \Gamma(\frac{d}{2(d-1)})}{\Gamma(\frac{1}{2(d-1)})} \cr 
&&+ \sum_{n=1}^{\infty} \Big\{  \frac{\Gamma(n+\frac{1}{2})}{\Gamma(n+1)} \frac {1}{nd+1}\frac{\Gamma(\frac{d(n+1)}{2(d-1)})}{\Gamma(\frac{(nd+1)}{2(d-1)})} 
 - \frac{1}{\sqrt{2d(d-1)}n}\Big\}  \nonumber \\
 &&+v^2 \sum_{n=0}^{\infty}   \frac{\Gamma(n+\frac{1}{2})}{\Gamma(n+1)} \left[  
 \frac{2 \Gamma[\frac{d(n+2)}{2(d-1)}]}{(d(n-1)+3)\Gamma[\frac{d(n-1)+3}{2(d-1)}]} \right. \nonumber \\
 &&- \left.  \frac{ \Gamma[\frac{d(n+4)-2}{2(d-1)}]}{{(d(n+1)+1)\Gamma[\frac{d(n+1)+1}{2(d-1)}]}}  -  \frac{ \Gamma[\frac{d(n+1)}{2(d-1)}]}{{(d(n-2)+3)\Gamma[\frac{d(n-2)+3}{2(d-1)}]}} \right] \cr 
 && \cr 
 && + \mathcal{O}(\epsilon). 
 \label{logeps}
\end{eqnarray}
Solving the above equation for $\epsilon$ and considering the leading order term we get,
\begin{eqnarray}
\label{eq:epsilon}
\epsilon = \mathcal{E}_{ent}^v e^{-\sqrt{\frac{d(d-1)}{2}}~\frac{l^{||}r_H}{R^2}},
\end{eqnarray}
where we have defined $\mathcal{E}^v_{ent}$ as,
\begin{eqnarray}
\mathcal{E}_{ent}^v &=& \frac{1}{d} \, \exp \Big[ \sqrt{2 d(d-1)}\bigg( \frac{\sqrt{\pi} \Gamma(\frac{d}{2(d-1)})}{\Gamma(\frac{1}{2(d-1)})}\cr 
&& + \sum_{n=1}^{\infty} \Big\{  \frac{\Gamma(n+\frac{1}{2})}{\Gamma(n+1)} \frac {1}{nd+1}\frac{\Gamma(\frac{d(n+1)}{2(d-1)})}{\Gamma(\frac{(nd+1)}{2(d-1)})} - \frac{1}{\sqrt{2d(d-1)}n}
 \Big\} \bigg) \nonumber \\
  &&+v^2 \sum_{n=0}^{\infty}   \frac{\Gamma(n+\frac{1}{2})}{\Gamma(n+1)} \Big[  
 \frac{2 \Gamma[\frac{d(n+2)}{2(d-1)}]}{(d(n-1)+3)\Gamma[\frac{d(n-1)+3}{2(d-1)}]}  \nonumber \\
 &&-  \frac{ \Gamma[\frac{d(n+4)-2}{2(d-1)}]}{{(d(n+1)+1)\Gamma[\frac{d(n+1)+1}{2(d-1)}]}}  -  \frac{ \Gamma[\frac{d(n+1)}{2(d-1)}]}{{(d(n-2)+3)\Gamma[\frac{d(n-2)+3}{2(d-1)}]}} \Big].
 \end{eqnarray} 
We have already checked that the finite part of the minimal area (\ref{noepsfinite}) behaves regularly when there was no $\epsilon$ correction.  However, as we consider $r_t^{||} = (1+ \epsilon) r_H$, we need to make sure that no new divergent term at the linear order in $\epsilon$ should further appear in the expression of finite area. To investigate the possibility of any such divergence at the linear order in $\epsilon$ and hence to regulate them appropriately, we rearrange terms in (\ref{approxA2rw1}),

 \begin{eqnarray}
 && \mathcal{A}^{||}_{\text{finite}} = 2 \frac{L^{d-2}}{R^{d-3}}  {r_t^{||}}^{d-2} \left[  \frac{l {r_t^{||}}}{2R^2} -  \frac{(d-1)\sqrt{\pi}\Gamma[\frac{d}{2(d-1)}]}{(d-2) \Gamma[\frac{1}{2(d-1)}]} + \sqrt{\frac{d-1}{2d^3}} Li_2(\alpha^d) \right. \cr 
 &&+    \sum_{n=1}^{\infty}  \alpha^{nd} \left\{ \frac{\Gamma(n+\frac{1}{2}) }{\Gamma(n+1)} \frac{1}{nd+1} \Big( \frac{d-1}{d(n-1)+2} \Big) \frac{\Gamma[\frac{d(n+1)}{2(d-1)}]}{\Gamma[\frac{nd+1}{2(d-1)}] }  - \sqrt{\frac{d-1}{2d^3}} \frac{1}{n^2} \right\} \cr 
 &&- \left. v^2\alpha^d \sum_{n=0}^{\infty} \alpha^{nd} \frac{\Gamma(n+\frac{1}{2})}{ \Gamma(n+1)}\left\{ \frac{\Gamma(\frac{d(n+1)}{2(d-1)})}{2(d-1)\Gamma(\frac{nd+1}{2(d-1)})} - \frac{\Gamma(\frac{d(n+2)}{2(d-1)})}{2(d-1)\Gamma(\frac{(n+1)d+1}{2(d-1)})} \right. \right. \cr 
 &&+ \left. \left.  \frac{2 \Gamma[\frac{d(n+2)}{2(d-1)}]}{(d(n-1)+3)\Gamma[\frac{d(n-1)+3}{2(d-1)}]} -  \frac{ \Gamma[\frac{d(n+4)-2}{2(d-1)}]}{{(d(n+1)+1)\Gamma[\frac{d(n+1)+1}{2(d-1)}]}}  -  \frac{ \Gamma[\frac{d(n+1)}{2(d-1)}]}{{(d(n-2)+3)\Gamma[\frac{d(n-2)+3}{2(d-1)}]}}\right\} \right] .\cr 
 && \cr
 &&
   \label{approxA2rw}
 \end{eqnarray}

In the above  expression we have used the series representation of the polylog function, $Li_2(z) = \sum_{n = 1}^{\infty} \frac{z^n}{n^2}$. 
Now we expand terms in the powers of $\epsilon$ and keep terms up to $\mathcal{O}(\epsilon)$\footnote{For small $\epsilon$,  $Li_2(\alpha^d) = \frac{\pi^2}{6}+ d \epsilon (-1+ \log(d\epsilon))+ \cdots $}.
  
  \begin{equation}
  \mathcal{A}^{||}_{\text{finite}} = \frac{L^{d-2}  l^{||}  r_H^{d-1}}{R^{d-1}} \left[ 1 +\frac{2R^2}{l^{||} r_H} \Big\{\Big( \mathcal{S}^{||}_0   -\epsilon \sqrt{\frac{d-1}{2d}}   \Big)+  v^2\Big( \mathcal{S}^{||}_1 +  \epsilon ~ \mathcal {S}^{||}_3 \Big) \Big\}\right] +\mathcal{O}(\epsilon^2), 
  \label{eq:a_finite}
 \end{equation}
where
\begin{eqnarray}
 && \mathcal{S}^{||}_3 =
 \sum_{n=0}^{\infty}   \frac{\Gamma(n+\frac{1}{2})}{\Gamma(n+1)} \left\{  
 \frac{2 \Gamma[\frac{d(n+2)}{2(d-1)}]}{(d(n-1)+3)\Gamma[\frac{d(n-1)+3}{2(d-1)}]} \right. \cr 
 && \qquad\qquad\qquad\quad - \left.  \frac{ \Gamma[\frac{d(n+4)-2}{2(d-1)}]}{{(d(n+1)+1)\Gamma[\frac{d(n+1)+1}{2(d-1)}]}}  -  \frac{ \Gamma[\frac{d(n+1)}{2(d-1)}]}{{(d(n-2)+3)\Gamma[\frac{d(n-2)+3}{2(d-1)}]}} \right\}\nonumber\\
 &&+ \sum_{n=0}^{\infty}  \frac{(nd+2)\Gamma(n+\frac{1}{2})}{ \Gamma(n+1)}\left\{ \frac{\Gamma(\frac{d(n+1)}{2(d-1)})}{2(d-1)\Gamma(\frac{nd+1}{2(d-1)})} - \frac{\Gamma(\frac{d(n+2)}{2(d-1)})}{2(d-1)\Gamma(\frac{(n+1)d+1}{2(d-1)})} \right.  \cr 
 &&+  \left.  \frac{2 \Gamma[\frac{d(n+2)}{2(d-1)}]}{(d(n-1)+3)\Gamma[\frac{d(n-1)+3}{2(d-1)}]} -  \frac{ \Gamma[\frac{d(n+4)-2}{2(d-1)}]}{{(d(n+1)+1)\Gamma[\frac{d(n+1)+1}{2(d-1)}]}}  -  \frac{ \Gamma[\frac{d(n+1)}{2(d-1)}]}{{(d(n-2)+3)\Gamma[\frac{d(n-2)+3}{2(d-1)}]}}\right\}. \cr 
 && \cr
 &&
\end{eqnarray}

 
Using the Ryu-Takayangi's prescription, once again we compute the holographic entanglement entropy for boosted plasma up to $\mathcal{O}({\epsilon})$,

\begin{eqnarray}
S^{||} &=& \frac{R^{d-1}}{4 G_N^{d+1}} \Big [ \frac{2}{d-2} {\Big(\frac{L}{\delta^{||}}\Big)}^{d-2} +
V_{d-1} {\Big(\frac{4 \pi T_{\text{boost}}}{d}\Big)}^{d-1} \{ 1+ v^2\Big(\frac{d-1}{2}\Big)\}  \cr 
&+& A_{d-2} {\Big(\frac{4 \pi T_{\text{boost}}}{d}\Big)}^{d-2} \left\{ \left(\mathcal{S}^{||}_0 -\epsilon \sqrt{\frac{d-1}{2d}}\right) + v^2 \Big(\mathcal{S}^{||}_1  + \frac{d-2}{2} \mathcal{S}{||}_0 + \epsilon \left( \mathcal{S}^{||}_3-(d-1)\sqrt{\frac{d-1}{2d}}\right)\Big) \right\} \Big] + \mathcal{O}(v^4). \cr 
&&
\label{entropyfinal2}
\end{eqnarray}

Comparing eqn (\ref{entropyfinal1})  with eqn (\ref{entropyfinal2}) we immediately observe that the term proportional to $V_{d-1}~ {T_{\text{boost}}}^{d-1} $ is exact in $\epsilon$ correction whereas the area dependent term $A_{d-2} ~{T_{\text{boost}}}^{d-2} $ gets modified in $\epsilon$.

\subsection{Holographic computation in perpendicular case}
In this section we carry out the holographic analysis of the entanglement entropy of boosted plasma keeping the direction of the boost and the orientation of the width $l$ perpendicular to each other. This is what we call $perpendicular$ case. The analysis in this section is qualitatively similar to the parallel case and thus to avoid repetition we mostly mention final results. To accomplish the holographic computation we recast the boosted bulk geometry in the following way,%
\begin{equation}
 ds^2 = \frac{r^2}{R^2}\left[ -dt^2 + 
 dx^2 +\gamma^2 \frac{r_H^d}{r^d}\left( dt+v dx\right)^2+dx_\perp^2+ d\vec{x}^2_{d-3}+\frac{R^4}{r^4}\frac{dr^2}{1-\frac{r_H^d}{r^d}}\right], 
 \label{boostedmetricperp}
\end{equation}
where we assume the thermal plasma is boosted along $x$ direction and the length of interest which is nothing but entangling width is along $x_\perp$ direction (orthogonal to $x$ direction).										

\begin{equation}
x_\perp\in\left[-\frac{l}{2},\frac{l}{2}\right]\quad x\in\left[-\frac{L}{2},\frac{L}{2}\right]; \quad x^i \in \left[-\frac{L}{2},\frac{L}{2}\right] \quad (i=1,2,\dots d-3),
\end{equation}

%
%
with $r' \equiv \frac{dr}{dx_\perp}$.  
The area functional reads as, 
\begin{align}
\label{eefunctionalperp}
 \mathcal{A}^{\perp} = \frac{L^{d-2}}{R^{d-2}} \int dr r^{d-3} {\left[ \bigg(\frac{r^2}{R^2} x_\perp'^2 
 + \frac{R^2}{r^2} \left(1-\frac{r_H^d}{r^d}\right)^{-1}\bigg) \bigg(r^2 + \gamma^2 v^2 \frac{r_h^d}{r^{d-2}}\bigg)\right]}^{\frac{1}{2}},
\end{align}
with ${x_\perp}' \equiv \frac{dx_\perp}{dr}$. 
Now using the appropriate boundary conditions
\begin{eqnarray}
 \lim_{x_\perp' \to \infty}  r=r_t^{\perp}, ~~~~~  \lim_{r \to \infty} x_\perp(r) = \pm \frac{l^{\perp}}{2}.
 \end{eqnarray}
we arrive at,
\begin{eqnarray}
\nonumber
&&  \frac{l^{\perp}}{2} =\frac{R^2}{r_t^{\perp}}  \sum_{n=0}^{\infty} \frac{\alpha^{nd}}{nd+1} \frac{\Gamma(n+\frac{1}{2})\Gamma(\frac{d(n+1)}{2(d-1)})}{\Gamma(n+1)\Gamma(\frac{(nd+1)}{2(d-1)})}\\ \nonumber
&&  +\frac{R^2  v^2 \alpha^d}{r_t^{\perp}}  \sum_{n=0}^{\infty} \alpha^{nd}  \frac{\Gamma(n+\frac{1}{2})}{\Gamma(n+1)} \Big\{  
\frac{ \Gamma[\frac{d(n+2)}{2(d-1)}]}{(d(n-1)+3)\Gamma[\frac{d(n-1)+3}{2(d-1)}]}    -  \frac{ \Gamma[\frac{d(n+1)}{2(d-1)}]}{{(d(n-2)+3)\Gamma[\frac{d(n-2)+3}{2(d-1)}]}} \Big\} + \mathcal{O}(v^4).
 \label{eqngeneral1perp}
 \end{eqnarray}
The extremized area function contains both infinite and finite contributions. Although the infinite contribution ${ \mathcal{A}^{\perp} }_{\text{infinite}} = \frac{2}{d-2} \frac{L^{d-2} R^{d-1} }{{\delta^\perp}^{d-2}}$ is same as the parallel case and ${\delta^\perp}^{d-2}$ is the UV cut-off, whereas the finite part modifies in a non-trivial way. 
Finally, we focus into the high temperature limit $(l T_{\text{boost}} >> 1)$ and in order to implement the limit we express the action functional in the following prescribed form,
\begin{eqnarray}
&&   {\mathcal{A}^{\perp}}_{\text{finite}} = 2 \frac{L^{d-2}}{R^{d-3}}  r_t^{d-2} \Big [ \frac{l r_t}{2R^2} -  \frac{(d-1)\sqrt{\pi}\Gamma[\frac{d}{2(d-1)}]}{(d-2) \Gamma[\frac{1}{2(d-1)}]}    \sum_{n=1}^{\infty} \frac{\Gamma(n+\frac{1}{2}) \alpha^{nd}}{\Gamma(n+1)} \frac{1}{nd+1} \Big( \frac{d-1}{d(n-1)+2} \Big) \frac{\Gamma[\frac{d(n+1)}{2(d-1)}]}{\Gamma[\frac{nd+1}{2(d-1)}] } \nonumber \\
&& - v^2 \alpha^d \sum_{n=0}^{\infty} \alpha^{nd}\frac{\Gamma(n+\frac{1}{2})}{ \Gamma(n+1)} \Big[\frac{\Gamma(\frac{nd+2)}{2(d-1)})}{2(d-1)\Gamma(\frac{(n-1)d+3}{2(d-1)})} - \frac{2\Gamma(\frac{d(n+2)}{2(d-1)})}{2(d-1)\Gamma(\frac{(n+1)d+1}{2(d-1)})} + \frac{ \Gamma[\frac{d(n+2)}{2(d-1)}]}{(d(n-1)+3)\Gamma[\frac{d(n-1)+3}{2(d-1)}]}  \nonumber \\
&&  + \frac{ \Gamma[\frac{d(n+1)}{2(d-1)}]}{{2(d-1)\Gamma[\frac{nd+1}{2(d-1)}]}}  -  \frac{ \Gamma[\frac{d(n+1)}{2(d-1)}]}{{(d(n-2)+3)\Gamma[\frac{d(n-2)+3}{2(d-1)}]}} \Big ] + \mathcal{O}(v^4).
\label{approxA2rw1perp}
\end{eqnarray}

Note that  infinite series present in (\ref{approxA2rw1perp}) does not give rise to any new divergent term. Now by imposing the limit $r_t \to r_H$ in (\ref{approxA2rw1perp}), one can obtain the leading behavior of the minimal area as, 
\begin{eqnarray}
\mathcal{A}_{\text{finite}} & = \frac{V_{d-1} r_H^{d-1}}{R^{d-1}} \Big[  1 + \frac{2 R^2}{l^{\perp} r_H} (\mathcal{S}^{\perp}_0 + v^2  \mathcal{S}^{\perp}_1)\Big] + \mathcal{O}(v^4),
\label{noepsfiniteperp}
\end{eqnarray}
where we have denoted the spatial volume of the rectangular strip as $V_{d-1} = l L^{d-2}$ and $\mathcal{S}_0$ and $\mathcal{S}_1$ can be expressed as,
\begin{eqnarray}
\mathcal{S}^{\perp}_0  &=& \Big(  -  \frac{(d-1)\sqrt{\pi}\Gamma[\frac{d}{2(d-1)}]}{(d-2) \Gamma[\frac{1}{2(d-1)}]}  +  \sum_{n=1}^{\infty} \frac{\Gamma(n+\frac{1}{2}) }{\Gamma(n+1)} \frac{1}{nd+1} \Big( \frac{d-1}{d(n-1)+2} \Big) \frac{\Gamma[\frac{d(n+1)}{2(d-1)}]}{\Gamma[\frac{nd+1}{2(d-1)}] }  \Big), \nonumber \\
\mathcal{S}^{\perp}_1 &=& -\sum_{n=0}^{\infty} \frac{\Gamma(n+\frac{1}{2})}{ \Gamma(n+1)} \Big[\frac{\Gamma(\frac{nd+2)}{2(d-1)})}{2(d-1)\Gamma(\frac{(n-1)d+3}{2(d-1)})} - \frac{2\Gamma(\frac{d(n+2)}{2(d-1)})}{2(d-1)\Gamma(\frac{(n+1)d+1}{2(d-1)})} + \frac{ \Gamma[\frac{d(n+2)}{2(d-1)}]}{(d(n-1)+3)\Gamma[\frac{d(n-1)+3}{2(d-1)}]}  \nonumber \\
&&  + \frac{ \Gamma[\frac{d(n+1)}{2(d-1)}]}{{2(d-1)\Gamma[\frac{nd+1}{2(d-1)}]}}  -  \frac{ \Gamma[\frac{d(n+1)}{2(d-1)}]}{{(d(n-2)+3)\Gamma[\frac{d(n-2)+3}{2(d-1)}]}} \Big ].
\label{perpcoeff}
\end{eqnarray}
Finally, holographic entanglement entropy of the strongly coupled boosted plasma living in the strip region with entangling width $l^{\perp}$ along $x^{\perp}$ from the perspective of rest frame observer,
\begin{eqnarray}
S^{\perp} = \frac{R^{d-1}}{4 G_N^{d+1}} \Big [ \frac{2}{d-2} {\Big(\frac{L}{\delta}\Big)}^{d-2} +
V_{d-1} {\Big(\frac{4 \pi T_{\text{boost}}}{d}\Big)}^{d-1} \{ 1+ v^2\Big(\frac{d-1}{2}\Big)\}  \nonumber \\
+ A_{d-2} {\Big(\frac{4 \pi T_{\text{boost}}}{d}\Big)}^{d-2} \{ \mathcal{S}^{\perp}_0 + v^2 \Big(\mathcal{S}^{\perp}_1  + \frac{d-2}{2} \mathcal{S}^{\perp}_0\Big) \} \Big] + \mathcal{O}(v^4), 
\label{entropyfinal1perp}
\end{eqnarray}
where $A_{d-2} = 2 L^{d-2}$ is the spatial area of the rectangular strip. Since the $\epsilon$ correction analysis in the perpendicular case would not lead to any qualitatively new result as compared to parallel case, we plan to hold off that analysis for time being. Here also we compute the central charge ${\mathcal{C}}^{\perp}_{v}$ and observe no quantitative change as such. This probably implies the number of degrees freedom in the boosted plasma at any instant is independent of the relative orientation of the static observer with respect to the boost direction. However, most interesting, this is not the case for holographic entanglement entropy which we discuss in the following. 

It is important to note that the expressions for holographic entanglement entropy of boosted fluid for both parallel and perpendicular case are formally similar and we verify this fact up to the first order approximation ($r^{||} \rightarrow r_h$ or $r^{\perp} \rightarrow r_h$). However, due to the difference between coefficients $\mathcal{S}^{||}_1$ and $\mathcal{S}^{\perp}_1$ we estimate the variation of entanglement entropy due to the change of orientation of entangling width with respect to the direction of boost. To quantify such difference we compute the following,
\begin{eqnarray}
S^{||} - S^{\perp} = \frac{R^{d-1}}{4 G^{d+1}_{N}} \Big[ A_{d-2}{\Big(\frac{4 \pi T{\text{boost}}}{d} \Big)}^{d-2} v^2 (\mathcal{S}^{||}_1 - \mathcal{S}^{\perp}_1)  \Big].
\end{eqnarray}
By using the definition of $\mathcal{S}^{||}_1$ and $\mathcal{S}^{\perp}_1$ as given in eqn (\ref{paracoeff}) and in eqn (\ref{perpcoeff}) respectively we compute $S^{||} - S^{\perp}$ for four dimensional boundary theory and it turns out as,
\begin{eqnarray}
S^{||}_{d=4} - S^{\perp}_{d=4} =  -0.272525\frac{R^{d-1}}{4 G^{d+1}_{N}} \Big[ A_{d-2}{\Big(\frac{4 \pi T{\text{boost}}}{d} \Big)}^{d-2} v^2   \Big].
\end{eqnarray}
It is evident from the above analysis that $S^{||}_{d=4} < S^{\perp}_{d=4}$. This is also true for boundary theories living on $d>4$ spacetime. Therefore we conclude that the holographic entanglement entropy associated to strongly coupled boosted plasma in  perpendicular case is higher than the same for the parallel case.  
\section{Two point correlator in a boosted plasma} \label{s2}

After having discussed the holographic entanglement entropy of the strongly coupled boosted plasma, here we explore another important non-local observable, equal time two-point correlation function, by using the geodesic approximation method. Holographic computation of two-point correlation function in Euclidean signature  was first introduced in \cite{Gubser, Witten}. Further generalization for studying two point function directly from the Minkowski signature at finite temperature was prescribed in \cite{SonStarinets, SonHerzog}.

Following \cite{BalasubramanianRoss, Banks, Saddlepnt1, Balasubramanian:2011ur}, the equal time two point correlators in the strongly coupled boundary theory can be realized as a weighted sum over all possible paths starting from a boundary configuration $(t,x)$ and ending at $(t,x')$ as follows 
\begin{equation}
\left\langle  \mathcal{O}(t, x)\mathcal{O}(t, x')\right\rangle=  \int \mathcal{D}\mathcal{P}\,\, e^{-\Delta L(\mathcal{P})}.
\label{pathint}
\end{equation}
Here, $\Delta$ is the conformal dimension of the operator $\mathcal{O}$  and $L(\mathcal{P})$ is the proper length of the path $\mathcal{P}$.  The conformal dimension of the boundary operator is related to the bulk theory as $\Delta =(d+\sqrt{d^2+4m^2R^2})/2$, where $m$ is the mass of the bulk primary scalar and $R$ is the radius of curvature in the AdS spacetime.

By using the saddle point approximation, the equation (\ref{pathint}) turns into the discrete summation over the geodesics as follows,

 \begin{equation}
\left\langle  \mathcal{O}(t, x)\mathcal{O}(t, x')\right\rangle= \sum \, e^{-\Delta L_{geodesic}},
\label{pathint2}
\end{equation}
where $L_{geodesic}$ is the magnitude of the geodesic linking $(t, x)$ and $(t,x')$. Due to the existence of the logarithmic divergence in $L_{geodesic}$, the regularized geodesic length can be defined with the assumption of the cutoff $r_b$ as 
\begin{equation}
L^{ren}_{geodesic}=L_{geodesic}-2 \ln r_b,
\label{rengdnclnght}
\end{equation}
with which one eventually gets the regularized two-point function,
 \begin{equation}
\left\langle  \mathcal{O}(t, x)\mathcal{O}(t, x')\right\rangle = e^{-\Delta L^{ren}_{geodesic}}.
\label{pathint2}
\end{equation}

With this formal connection between the two point correletor and the geodesic length approximation method we proceed in the next section to discuss the holographic computation of two point equal time correletor  for the  strongly coupled \textit{boosted} large $N$ plasma at finite temperature.

\subsection{Holographic derivation}    
Here we compute the two-point correlation function $$\langle \mathcal{O}(t, -\frac{l}{2}, \vec{x}_{d-2}=\mathbf{0})\mathcal{O}(t, \frac{l}{2}, \vec{x}_{d-2}=\mathbf{0}) \rangle \equiv  \langle \mathcal{O}(t,-l/2)\mathcal{O}(t,l/2) \rangle $$ of scalar primary operators. The relevant part of the metric~\eqref{boostedmetric} to compute the geodesic length connecting the points $ (t, -\frac{l}{2}, \vec{x}_{d-2}=\mathbf{0})$ and $(t, \frac{l}{2}, \vec{x}_{d-2}=\mathbf{0})$ would be
\begin{equation}
\label{eqn11}
ds^2=r^2\left(1+\gamma^2\frac{r_H^d}{r^d}v^2\right) dx^2+\frac{1}{r^2\left(1-\frac{r_H^d}{r^d}\right)}dr^2,
\end{equation}
where we have fixed the AdS radius $R = 1$ for the sake of simplicity. Note that the present choice of spacelike boundary interval of width $l$ is exactly similar to the choice of entangling width $l^{||}$ as mentioned in the parallel case. It is straightforward to check that the analogue of choosing a spacelike interval in the boundary analogous to $l^{\perp}$ as given in perpendicular case reproduces the holographic computation of two point correlators in the unboosted plasma. This is intuitively expected as such spacelike interval is oriented in an entirely orthogonal way with respect to boost direction. To proceed further, we take the affine parameter to be the geodesic proper length $s$ and write the spacelike geodesic equations 
\begin{align}
\dot{x}=& \frac{dx}{ds} = \frac{r_t}{r^2} \frac{1}{1+\gamma^2\frac{r_H^d}{r^d }v^2},\label{xgeoeqn}\\
\dot{r}=& \frac{dr}{ds} = \pm r \sqrt{\left(1- \frac{r_t^2}{r^2}\frac{1}{1+\gamma^2\frac{r_H^d}{r^d }v^2}\right)\left(1-\frac{r_H^d}{r^d}\right)}
\label{rgeoeqn},
\end{align}
where the two branches of spacelike geodesic join smoothly at $r=r_t$. The branch starting from $(r\to \infty,x=l/2)$ and ending at $(r_t, x=0)$ is called the positive branch, whereas the negative branch starts from $(r\to \infty,x=-l/2)$ and ends at $(r_t, x=0)$.  Looking at (\ref{xgeoeqn}) and (\ref{rgeoeqn}) for the positive branch, one  arrives at
\begin{equation}
\frac{dr}{dx}= \frac{r^3}{r_t}\left(1+\gamma^2\frac{r_H^d}{r^d }v^2\right) \sqrt{\left(1-\frac{r_H^d}{r^d }\right)\left(1- \frac{1}{r^2}\frac{r_t^2}{1+\gamma^2\frac{r_H^d}{r^d }v^2}\right)}.
\label{rxgeoeqn}
\end{equation}
Together with the appropriate boundary conditions, one can solve the equation (\ref{rxgeoeqn}) as,
\begin{equation}
\begin{aligned}
\frac{l}{2}=&r_t\int_{r_t}^{\infty}\frac{ dr}{r^3 (1+\gamma^2\frac{r_H^d}{r^d }v^2) \sqrt{\left(1- \frac{1}{r^2}\frac{r_t^2}{1+\gamma^2\frac{r_H^d}{r^d }v^2}\right)}} \left(1-\frac{r_H^d}{r^d}\right)^{-1/2}\\
=& \frac{1}{r_t}\int_{0}^{1}\frac{u du}{\left(1+\gamma^2v^2\frac{r_H^d}{r_t^d }u^d\right) \sqrt{1- \frac{u^2}{1+\gamma^2v^2\frac{r_H^d}{r_t^d }u^d} }} \left(1-\frac{r_H^d}{r_t^d}u^d\right)^{-1/2}. 
\label{leqn}
\end{aligned}
\end{equation}
Following the similar consideration as described in the last section, we perform a perturbative expansion of the integrand in (\ref{leqn}) with respect to the boost parameter $v$ and keep all the terms up to $\mathcal{O}(v^2)$.  Now, by integrating the expanded version of (\ref{leqn}) we get,
\begin{equation}
\begin{aligned}
\frac{l}{2}=& \frac{1}{r_t}\int_{0}^{1}du\Bigg[\frac{u}{ \sqrt{1- u^2}}+v^2 \frac{u^{d+1}(-2+u^2)}{2(1- u^2)^{\frac{3}{2}}}(\frac{r_H}{r_t})^{d}\Bigg]\sum_{n=0}^\infty\frac{\Gamma\left[n+\frac{1}{2}\right]}{\sqrt{\pi } \Gamma[n+1]}\left(\frac{r_H}{r_t}\right)^{nd}u^{nd}\\
=&\frac{1}{2 r_t}\sum_{n=0}^\infty\frac{\Gamma\left[n+\frac{1}{2}\right]}{ \Gamma[n+1]}\Bigg\{\frac{\Gamma\left[\frac{nd+2}{2}\right]}{\Gamma\left[\frac{nd+3}{2}\right]}+v^2\frac{(n+1)d}{(n+1)d+1}\frac{\Gamma\left[\frac{(n+1)d+2}{2}\right]}{\Gamma\left[\frac{(n+1)d+1}{2}\right]}(\frac{r_H}{r_t})^{d}\Bigg\}\left(\frac{r_H}{r_t}\right)^{nd}.
\label{expansionlv}
\end{aligned}
\end{equation}
Notice that the large $n$ behavior of the $v$-independent part of the series is $\sim \frac{1}{n}\left(r_H/r_t\right)^{nd}$ and it converges as the condition $r_H/r_t<1$ is maintained. Moreover, the $v$ dependent contribution behaves as $\sim \left(r_H/r_t\right)^{nd}$ which is also convergent as $r_H/r_t<1$. However, in the high temperature regime both of these terms give rise to divergences.

After obtaining a relation between $r_t$ and $l$, we now compute the geodesic length by using~\eqref{rgeoeqn} as follows
\begin{equation}
\begin{aligned}
{L}=&2 \int_{r_t}^{\infty}\frac{ dr}{r \sqrt{1- \frac{r_t^2}{r^2}\frac{1}{1+\gamma^2\frac{r_H^d}{r^d}v^2}}} \left(1-\frac{r_H^d}{r^d}\right)^{-1/2}\\&
=2 \int_{r_t/r_b}^{1}\frac{ du}{u \sqrt{1- \frac{u^2}{1+\gamma^2v^2\frac{r_H^d}{r_t^d}u^d}}} \left(1-\frac{r_H^d}{r_t^d}u^d\right)^{-1/2}. 
\label{renleqn}
\end{aligned}
\end{equation}   
Observe that taking care of the two branches of the geodesic results in setting the factor of 2 in front of the integral. Expanding \eqref{renleqn} up to second order in $v$, one finds the regularized geodesic length to be
\begin{equation}
{L}_{ren}= 2\int_{r_t/r_b}^{1}du\Bigg[\frac{1}{u \sqrt{1- u^2}}-v^2 \frac{u^{d+1}}{2(1- u^2)^{\frac{3}{2}}}(\frac{r_H}{r_t})^{d}\Bigg]\sum_{n=0}^\infty\frac{\Gamma\left[n+\frac{1}{2}\right]}{\sqrt{\pi } \Gamma[n+1]}\left(\frac{r_H}{r_t}\right)^{nd}u^{nd}-2\ln r_b,
\end{equation}
where we have introduced a cutoff $r_b$ to remove the log divergence in $L$. After evaluating the integral, one ultimately gets
\begin{equation} 
{L}_{ren}=2\ln\left(\frac{2}{r_t}\right)+  \sum_{n=1}^\infty\frac{\Gamma\left[n+\frac{1}{2}\right]\Gamma\left[\frac{nd}{2}\right]}{ \Gamma[n+1]\Gamma\left[\frac{nd+1}{2}\right]}\left(\frac{r_H}{r_t}\right)^{nd}+v^2\sum_{n=0}^\infty\frac{\Gamma\left[n+\frac{1}{2}\right]\Gamma\left[\frac{(n+1)d+2}{2}\right]}{ \Gamma[n+1]\Gamma\left[\frac{(n+1)d+1}{2}\right]}\left(\frac{r_H}{r_t}\right)^{(n+1)d}.
\label{renfineqn}
\end{equation}
Again considering the large $n$ behavior of the summands that appear in the above expression one can show that $L_{ren}$ is well defined when $r_H/r_t<1$.
In what follows, by using (\ref{expansionlv}), we solve $r_t$ in the in the high temperature limit and study the two-point correlator. 

\subsection{High temperature behavior of two-point function}
In this section we compute the two point correlators as $T_{\text{boost}} l \gg 1$. In the dual bulk theory, we follow the geodesic approximation method keeping the fact in mind that  the near horizon part of the geodesic contributes to the leading order in the computation. 
\begin{equation}
\begin{aligned}
{L}_{ren}=& 2\ln\left(\frac{2}{r_t}\right)+\sum_{n=1}^\infty\left(\frac{n d+1}{nd}\right)\frac{\Gamma\left[n+\frac{1}{2}\right]\Gamma\left[\frac{nd+2}{2}\right]}{ \Gamma[n+1]\Gamma\left[\frac{nd+3}{2}\right]}\left(\frac{r_H}{r_t}\right)^{nd}\\
& \hskip 1.6 cm+v^2\sum_{n=0}^\infty\frac{\Gamma\left[n+\frac{1}{2}\right]\Gamma\left[\frac{(n+1)d+2}{2}\right]}{ \Gamma[n+1]\Gamma\left[\frac{(n+1)d+1}{2}\right]}\left(\frac{r_H}{r_t}\right)^{(n+1)d}.
\label{hightmprappr}
\end{aligned}  
\end{equation}
Once we implement the high temperature limit ($r_t \to r_h$), we observe the appearance of divergence in the above expression (\ref{hightmprappr}) of geodesic length. To bypass such divergence we re-write ${L}_{ren}$  by using  (\ref{expansionlv}), in the following way,
\begin{equation}
\begin{aligned}
{L}_{ren}=& 2\ln\left(\frac{2}{r_t}\right)+\left(r_t l -2\right)+\sum_{n=1}^\infty\left(\frac{1}{nd}\right)\frac{\Gamma\left[n+\frac{1}{2}\right]\Gamma\left[\frac{nd+2}{2}\right]}{ \Gamma[n+1]\Gamma\left[\frac{nd+3}{2}\right]}\left(\frac{r_H}{r_t}\right)^{nd}\\
& +v^2\sum_{n=0}^\infty \left(\frac{r_H}{r_t}\right)^{(n+1)d} \frac{\Gamma(n+\frac{1}{2})}{\Gamma(n+1)} \left[ \frac{\Gamma\left(\frac{(n+1)d+2}{2}\right)}{\Gamma\left(\frac{(n+1)d+1}{2}\right)} - 2\frac{\Gamma\left(\frac{(n+1)d+2}{2}\right)}{\Gamma\left(\frac{(n+1)d+2}{2}\right)} + \frac{\Gamma\left(\frac{(n+1)d+4}{2}\right)}{\Gamma\left(\frac{(n+1)d+3}{2}\right)}\right].
\end{aligned} \label{highrenl}
\end{equation} 
Note that under high temperature limit, the $v$ independent terms present in both (\ref{expansionlv}) and (\ref{renfineqn}) give rise to same type of divergences. However as we combine them in (\ref{highrenl}), those divergences cancel each other and ${L}_{ren}$ remains finite. Similarly, it is straightforward to check that the divergences related to high temperature limit present in the $v$ dependent terms in (\ref{highrenl}) also get nicely cancelled. 
\begin{eqnarray}
\frac{1}{\sqrt{n}}\left(\frac{nd}{2}\right)^{1/2} -2 \frac{1}{\sqrt{n}}\left(\frac{nd}{2}\right)^{1/2} + \frac{1}{\sqrt{n}}\left(\frac{nd}{2}\right)^{1/2} = 0
\end{eqnarray} 
As a first approximation, by using $r_t \sim r_H$ in the high temperature limit, geodesic length reads as,
\begin{equation}
\begin{aligned}
{L}_{ren}\approx & 2\ln\left(\frac{2}{r_H}\right)+\left(r_H l -2\right)+\sum_{n=1}^\infty\left(\frac{1}{nd}\right)\frac{\Gamma\left[n+\frac{1}{2}\right]\Gamma\left[\frac{nd+2}{2}\right]}{ \Gamma[n+1]\Gamma\left[\frac{nd+3}{2}\right]}\\& \hskip 3.8 cm+v^2\sum_{n=0}^\infty \frac{\Gamma(n+\frac{1}{2})}{\Gamma(n+1)} \left[ \frac{\Gamma\left(\frac{(n+1)d+2}{2}\right)}{\Gamma\left(\frac{(n+1)d+1}{2}\right)} - 2\frac{\Gamma\left(\frac{(n+1)d+2}{2}\right)}{\Gamma\left(\frac{(n+1)d+2}{2}\right)} + \frac{\Gamma\left(\frac{(n+1)d+4}{2}\right)}{\Gamma\left(\frac{(n+1)d+3}{2}\right)}\right].
\label{highleadcnt}
\end{aligned}
\end{equation}
Consequently, by inserting (\ref{highleadcnt}) into (\ref{pathint2}), one obtains the two-point correlation function as,
\begin{equation}
\langle \mathcal{O}(t,-l/2)\mathcal{O}(t,l/2) \rangle \approx \mathcal{C}_{d,\Delta, v}~ r_H^{2 \Delta}~ e^{-\Delta r_H l},
\label{twpntfnc}
\end{equation}
where
\begin{equation}
\begin{aligned}
 \mathcal{C}_{d,\Delta, v}=& {\left(\frac{1}{4} \exp\left({2-\sum_{n=1}^\infty\left(\frac{1}{nd}\right)\frac{\Gamma\left[n+\frac{1}{2}\right]\Gamma\left[\frac{nd+2}{2}\right]}{ \Gamma[n+1]\Gamma\left[\frac{nd+3}{2}\right]} }\right) \right)}^{\Delta} \\ \nonumber
 & \times\exp\left( - v^2 \Delta \sum_{n=0}^\infty \frac{\Gamma(n+\frac{1}{2})}{\Gamma(n+1)} \left[ \frac{\Gamma\left(\frac{(n+1)d+2}{2}\right)}{\Gamma\left(\frac{(n+1)d+1}{2}\right)}  - 2\frac{\Gamma\left(\frac{(n+1)d+2}{2}\right)}{\Gamma\left(\frac{(n+1)d+2}{2}\right)} + \frac{\Gamma\left(\frac{(n+1)d+4}{2}\right)}{\Gamma\left(\frac{(n+1)d+3}{2}\right)}\right]   \right).
\end{aligned}
\end{equation}

Finally we express the two point correlators in terms of the boundary parameters, viz.  $T_{\text{boost}}$ and the boost velocity $v$. 
\begin{equation}
\begin{aligned}
\langle \mathcal{O}(t, x)\mathcal{O}(t, x') \rangle\approx & \mathcal{C}_{d,\Delta} \left(\frac{4 \pi  T_{boost}}{d}\right)^{2 \Delta} e^{- 4\pi\Delta  T_{boost} |x-x'|/d}\bigg[1+ v^2 \Delta \bigg(1 -\frac{2\pi T_{boost}|x-x'|}{d}\\&- \sum_{n=0}^\infty  \frac{\Gamma(n+\frac{1}{2})}{\Gamma(n+1)} \left[ \frac{\Gamma\left(\frac{(n+1)d+2}{2}\right)}{\Gamma\left(\frac{(n+1)d+1}{2}\right)}  - 2\frac{\Gamma\left(\frac{(n+1)d+2}{2}\right)}{\Gamma\left(\frac{(n+1)d+2}{2}\right)} + \frac{\Gamma\left(\frac{(n+1)d+4}{2}\right)}{\Gamma\left(\frac{(n+1)d+3}{2}\right)}\right]  \bigg)\bigg],
\end{aligned}
\end{equation}
where $\mathcal{C}_{d,\Delta} =\lim_{v\to 0} \mathcal{C}_{d,\Delta,v}$. 
It is evident from the above expression that at the high temperature limit, the leading term in the two point correlator is exponentially decaying. The sub-leading contribution is solely due to the presence of the boost parameter in the theory.  Moreover, one may further generalize the above computation of two point correlator by using the relation $r_t = (1+ \epsilon) r_H$  to see the rate of approach of $r_c$ towards $r_H$ and its implication over the two point correlators. 

\section{Conclusion}\label{s3}

In this work, we have explored the behavior of non local observables for strongly coupled large $N$, thermal plasma where the boost is given along any one of the spatial boundary coordinates (say $x$). In particular, using the boosted AdS Schwarzschild blackhole background (\ref{boostedmetric}) as the dual bulk theory, we have holographically computed the entanglement entropy of a strip region in the boundary theory. In this computation, we keep both entangling width of the strip region and the boost, aligned along the same spatial boundary direction as well as orthogonal to each other. In both cases, we have explicitly computed the modification of both leading and sub-leading terms of the finite contribution to the entanglement entropy up to the quadratic order in the boost parameter. We have observed that the holographic entanglement entropy of the strongly coupled boosted plasma in perpendicular case is always higher than that the parallel  case. In order to provide an explanation of this difference in two cases for HEE, we emphasize that by boosting the boundary plasma we essentially invoke the breaking of rotational symmetry in the boundary theory. Therefore the observation made by static observer in both cases differs accordingly as the parallel case is connected to the perpendicular case by a mere rotation and again the rotational symmetry is broken.

We have also computed the two point correlators of the strongly coupled plasma by using the geodesic approximation method. We are interested in the geodesic which connects two points in the boundary theory specified as $ (t, -\frac{l}{2}, \vec{x}_{d-2}=\mathbf{0})$ and $(t, \frac{l}{2}, \vec{x}_{d-2}=\mathbf{0})$. In this analysis we observe that the leading contribution to two point correlators remains exact in boost parameter and behaves as a exponentially decaying function of the width $l$, whereas the next sub-leading term is proportional to the quadratic power of the boost parameter. Here we notice that the perpendicular case does not lead to a modification of two point correlator due to the presence of boost in the theory.

In our analysis, the only length scale available is specified by a characteristic length $l$, inverse of which automatically defines a characteristic temperature scale in the theory. It turns out that one way to achieve the analytical result for non-local boundary observable is to consider the temperature of the plasma to be very high as compared to the characteristic temperature scale. It is necessary to parameterize the boundary non-local observables in terms of boundary entities measured from a specific frame of reference. Here, we prefer to present our result from the point of view of rest frame observer. Hence, we express the final form of the boundary non-local observable in terms of $T_{\text{boost}}$, $l$ and the boost parameter $v$.  Using the definition of the holographic $c$-function given in \cite{Myers:2012ed} we have also shown that the boost enhances the degrees of freedom as compared to that in the un-boosted case. 

Achieving modification of boundary observables due to arbitrary value of boost parameter $v$ is beyond the scope of analytical technics. Instead, we assume perturbative expansion in the power of $v$ and compute the results up to the quadratic power in $v$. It turns out the boost dependent corrections present in both EE and two point correlators do not bring in any new divergence at the high temperature limit.

It would be interesting to check how the holographic entanglement entropy of a strip region in the boundary  plasma varies with the boost $v$, where the direction of boost and the alignment of the entangling width $l$ are perpendicular to each other. Like wise one can compute the two point correlator using a geodesic connecting the points $ (t, 0, x_{1} = -L, \vec{x}_{d-3}=\mathbf{0})$ and $(t, 0, x_{1} = L, \vec{x}_{d-3}=\mathbf{0})$. Further, one can study the holographic entanglement entropy of a spherical region in the boosted plasma. In that case due to the present of boost along $x$ axis in the boundary, spherical symmetry of the entangling region will be modified accordingly.  In \cite{Pedraza:2014moa, DiNunno:2017obv}, the authors have studied low temperature behavior of various non-local observables in the boundary theory by actually doing the computation in the dual bulk theory described by a black hole that is boosted along the holographic direction. The dual boundary theory is a thermal plasma that is expanding and cooling down. It would interesting to study the high temperature behavior of non local observables in such theories. As an immediate generalization, it would be highly interesting to introduce dissipation in the boosted thermal plasma and study the behavior of non local observables within the high temperature limit. 

It is natural to ask whether our analysis of non-local observables for a boosted thermal plasma in the high temperature limit can be extended for arbitrary values of  boost parameter. Such analysis requires a systematic series expansion of the integrands given in eqn (\ref{eqn1}) and also in eqn (\ref{finalfinite}) in both temperature and boost parameter. Such double series expansion has already been presented in \cite {Kundu:2016dyk, Chakrabortty:2020ptb}. In this regard, we mention that although in principle, such double series expansion is also possible in our analysis of holographic entanglement entropy in both perpendicular and parallel cases, it is very hard to provide an analytical proof of the non-existence of unphysical divergence at all orders of boost parameter in such expansion. We check that if we generalize our analysis beyond quadratic order, no contribution appears from odd power of boost. If we keep our analysis limited up to quartic power of boost, the cancellation of divergence similar to quadratic order of boost nicely works. However, it would be really worth to guess a closed analytical form of entanglement entropy at high temperature for arbitrary value of boost by investigating contributions coming from beyond the quartic power. We thank the anonymous referee for suggesting to highlight this issue. 

\section{\label{ackno} Acknowledgments}

We would like to thank Mohsen Alishahiha, Arjun Bagchi, Nabamita Banerjee, Rudranil Basu, Jyotirmoy Bhattacharya, Umut Gursoy, Nabil Iqbal, R Loganayagam, Sudipta Mukherjee,  Bala Sathiapalan, Nemani Suryanarayan,  for useful suggestions. SC is partially supported by ISIRD grant 9-252/2016/IITRPR/708. The works of S.D. and E.K. are supported by the TUBITAK Grant No. 119F241.

\appendix

\section{Hawking temperature for boosted black brane }
 We start with the boosted black brane metric written in AdS-Schwarzschild coordinate
 \bea
 ds^2 = -\frac{r^2}{R^2}\left(1-\gamma^2\frac{r_H^d}{r^d}\right)dt^2 + 2 v \gamma^2 \frac{r_H^d}{r^{d-2} R^2} dt\, dx+ \frac{r^2}{R^2}\left(1+ v^2\gamma^2\frac{r_H^d}{r^d}\right) dx^2 +  \frac{r^2}{R^2} d\vec{x}^2_{d-2} + \frac{R^2}{r^2}\frac{dr^2}{1-\frac{r_H^d}{r^d}}. \cr 
 \eea
 The hypersurface $r=r_H$ is a null hypersurface. The normal on this hypersurface is
 \bea
  n_\alpha = \{0,\dots, 0,1\}.
 \eea
  We can show that 
  \bea
   \left. g^{\mu\nu} n_\mu n_\nu \right|_{r=r_H} =0.
  \eea
  The metric is symmetric under time translation, therefore $\xi_t= \{1,0,\dots,0\}$ is a Killing vector. Similarly, $\xi_x = \{0,1,\dots,0\}$ is also a Killing vector. One can also check that these vectors satisfy the Killing equations
  \bea
   \xi^\alpha \partial_\alpha g_{\mu\nu} + g_{\mu \alpha} \partial_\nu \xi^\alpha + g_{\nu \alpha} \partial_\mu \xi^\alpha =0.
  \eea
  We now consider observers moving in $x$ direction with an arbitrary, but uniform velocity $dx/dt = \beta$. They move with a four-velocity
  \bea
   u_s^\alpha = \Omega (\xi_t^\alpha + \beta \xi_x^\alpha ).
  \eea
Note that $\xi_t^\alpha + \beta \xi_x^\alpha $ is a Killing vector. The normalization factor $\Omega$ is given by
  \bea
   \Omega^{-2}&=& - g_{\mu\nu} (\xi_t^\mu + \beta \xi_x^\mu )(\xi_t^\nu + \beta \xi_x^\nu ) \cr 
    &=& -g_{tt} - 2\beta g_{tx}- \beta^2 g_{xx} \cr 
    &=& -g_{xx} (\beta^2-2b\beta+ g_{tt}/g_{xx}),
  \eea
  where $b= -g_{tx}/g_{xx}$. Outside the event horizon the vector $\xi_t^\alpha + \beta \xi_x^\alpha $ must be timelike, and on the horizon it must be null. The condition $\Omega^{-2}>0$ gives rise to the following requirement on the velocity of the observer:
  \bea
   \beta_- < \beta < \beta_+,
  \eea
  where $\beta_\pm = b \pm \sqrt{b^2-g_{tt}/g_{xx}}$. At the situation $\beta_-=\beta_+$ which implies $\beta =b$; the observer is forced to move with a velocity equal to $b$. This occurs when 
  \bea
   b^2- \frac{g_{tt}}{g_{xx}} =0.
  \eea
  For example in $d=4$ this condition becomes 
  \bea
   r^4 - r_H^4 (1 - v^2) \gamma^2 = 0.
  \eea
  The largest solution is $r=r_+=r_H$. The Killing vector $\xi_t^\alpha + \beta \xi_x^\alpha $ becomes null at $r=r_+=r_H$ which we identify with black brane's event horizon.
  
  To confirm that $r=r_H$ is truly the event horizon, we use the property that in a stationary spacetime, the event horizon is also an apparent horizon -- a surface of zero expansion for a congruence of outgoing null geodesics orthogonal to the surface. The event horizon must therefore be a null, stationary surface. We have already shown in the beginning that the surface is null. As the surface is independent of $t$ the surface is stationary.
  
  At $r=r_+=r_H$, one gets
  \bea
   b(r_+)= -v,
  \eea
  and the null Killing vector 
  \bea
    \xi^\alpha = \xi_t^\alpha + b(r) \xi_x^\alpha.
  \eea
  The surface gravity, $\kappa$ is defined as
  \bea
  \label{sgeq}
   \nabla_\alpha(-\xi^\beta \xi_\beta) = 2 \kappa \xi_\alpha.
  \eea
  As $r=r_H$ is an null hypersurface, 
  \bea
   \xi_\alpha = f(x^\mu) n_\alpha \qquad \text{with} \qquad f(x^\mu)= \frac{1}{\sqrt{1+v^2 \gamma^2 \frac{r_H^d}{r^d}}}.
  \eea
  At $r=r_H$
  \bea
   \xi_\alpha = \{0,0,\dots,0,1/\gamma\}.
  \eea
  Now
  \bea
   \left. \nabla_\alpha(-\xi^\beta \xi_\beta) \right|_{r=r_H} = \left. \partial_\alpha(-\xi^\beta \xi_\beta) \right|_{r=r_H} = \{ 0,0,\dots,0,\frac{d  r_H}{\gamma^2 R^2}  \}.
  \eea
  Therefore from \eqref{sgeq} we find
  \bea
   \kappa = \frac{d r_H}{2 \gamma R^2},
  \eea
  and the Hawking temperature
  \bea
  \label{hawkingtemp}
   T_H = \frac{\kappa}{2\pi}= \frac{d}{4\pi R^2}\,\frac{r_H}{\gamma}.
  \eea


\begin{thebibliography}{999}

\bibitem{Adcox:2004mh} 
  K.~Adcox {\it et al.} [PHENIX Collaboration],
  ``Formation of dense partonic matter in relativistic nucleus-nucleus collisions at RHIC: Experimental evaluation by the PHENIX collaboration,''
  Nucl.\ Phys.\ A {\bf 757}, 184 (2005)

  
  \bibitem{Adams:2005dq} 
  J.~Adams {\it et al.} [STAR Collaboration],
  ``Experimental and theoretical challenges in the search for the quark gluon plasma: The STAR Collaboration's critical assessment of the evidence from RHIC collisions,''
  Nucl.\ Phys.\ A {\bf 757}, 102 (2005)

  
  
  \bibitem{Back:2004je} 
  B.~B.~Back {\it et al.},
  ``The PHOBOS perspective on discoveries at RHIC,''
  Nucl.\ Phys.\ A {\bf 757}, 28 (2005)
  
  \bibitem{Baier:1996kr} 
  R.~Baier, Y.~L.~Dokshitzer, A.~H.~Mueller, S.~Peigne and D.~Schiff,
  ``Radiative energy loss of high-energy quarks and gluons in a finite volume quark - gluon plasma,''
  Nucl.\ Phys.\ B {\bf 483}, 291 (1997)


\bibitem{Eskola:2004cr} 
  K.~J.~Eskola, H.~Honkanen, C.~A.~Salgado and U.~A.~Wiedemann,
  ``The Fragility of high-p(T) hadron spectra as a hard probe,''
  Nucl.\ Phys.\ A {\bf 747}, 511 (2005)
  
\bibitem{Shuryak:2008eq} 
  E.~Shuryak,
  ``Physics of Strongly coupled Quark-Gluon Plasma,''
  Prog.\ Part.\ Nucl.\ Phys.\  {\bf 62}, 48 (2009).
  
  \bibitem{Shuryak:2014zxa} 
  E.~Shuryak,
  ``Strongly coupled quark-gluon plasma in heavy ion collisions,''
  Rev.\ Mod.\ Phys.\  {\bf 89}, 035001 (2017)
  
  \bibitem{Panero} 
  M.~Panero,
  ``Thermodynamics of the QCD plasma and the large-N limit,''
  Phys.\ Rev.\ Lett.\  {\bf 103}, 232001 (2009).
 
\bibitem{Kozcaz} 
  C.~P.~Herzog, A.~Karch, P.~Kovtun, C.~Kozcaz and L.~G.~Yaffe, ``Energy loss of a heavy quark moving through N=4 supersymmetric Yang-Mills plasma,'' JHEP {\bf 0607}, 013 (2006).

\bibitem{Maldacena:1997re} 
  J.~M.~Maldacena,
  ``The Large N limit of superconformal field theories and supergravity,''
  Int.\ J.\ Theor.\ Phys.\  {\bf 38}, 1113 (1999).
  
  \bibitem{Gubser} 
  S.~S.~Gubser, I.~R.~Klebanov and A.~M.~Polyakov,``Gauge theory correlators from noncritical string theory,'' Phys.\ Lett.\ B {\bf 428}, 105 (1998).
  
  \bibitem{Witten} 
  E.~Witten, ``Anti-de Sitter space and holography,'' Adv.\ Theor.\ Math.\ Phys.\  {\bf 2}, 253 (1998).
  
  \bibitem{Witten:1998zw} 
  E.~Witten,
  ``Anti-de Sitter space, thermal phase transition, and confinement in gauge theories,''
  Adv.\ Theor.\ Math.\ Phys.\  {\bf 2}, 505 (1998)
  
  \bibitem{Fischler:2012ca} 
  W.~Fischler and S.~Kundu,
  ``Strongly Coupled Gauge Theories: High and Low Temperature Behavior of Non-local Observables,''
  JHEP {\bf 1305}, 098 (2013)
  
  
  

 

  \bibitem{Myers_Ent} 
  D.~D.~Blanco, H.~Casini, L.~Y.~Hung and R.~C.~Myers,
  ``Relative Entropy and Holography,''
  JHEP {\bf 1308}, 060 (2013).
  
  
  
  
  \bibitem{Bhattacharyya:2008jc} 
  N.~Banerjee, J.~Bhattacharya, S.~Bhattacharyya, S.~Dutta, R.~Loganayagam and P.~Surowka,
  ``Hydrodynamics from charged black branes,''
  JHEP {\bf 1101}, 094 (2011)



  
  \bibitem{Hubeny:2009rc} 
  V.~E.~Hubeny, D.~Marolf and M.~Rangamani,
  ``Hawking radiation from AdS black holes,''
  Class.\ Quant.\ Grav.\  {\bf 27}, 095018 (2010)
  
  
  
  \bibitem{Horowitz:1996ay} 
  G.~T.~Horowitz, J.~M.~Maldacena and A.~Strominger,
  ``Nonextremal black hole microstates and U duality,''
  Phys.\ Lett.\ B {\bf 383}, 151 (1996)

  
  
  
\bibitem{Mozaffara:2016iwm} 
  M.~Reza Mohammadi Mozaffar, A.~Mollabashi and F.~Omidi,
  ``Non-local Probes in Holographic Theories with Momentum Relaxation,''
  JHEP {\bf 1610}, 135 (2016)
  doi:10.1007/JHEP10(2016)135
  [arXiv:1608.08781 [hep-th]].


\bibitem{Karar:2018hvy} 
  S.~Karar, S.~Gangopadhyay and A.~S.~Majumdar,
  ``Holographic complexity of ?black? non-susy D3-brane and the high temperature limit,''
  Int.\ J.\ Mod.\ Phys.\ A {\bf 34}, no. 01, 1950003 (2019)
  doi:10.1142/S0217751X19500039
  [arXiv:1804.00615 [hep-th]].
  
  \bibitem{Mishra:2016yor} 
  R.~Mishra and H.~Singh,
  ``Entanglement asymmetry for boosted black branes and the bound,''
  Int.\ J.\ Mod.\ Phys.\ A {\bf 32}, no. 16, 1750091 (2017)
  
  \bibitem{Karar:2019wjb} 
  S.~Karar, R.~Mishra and S.~Gangopadhyay,
  ``Holographic complexity of boosted black brane and Fisher information,''
  Phys.\ Rev.\ D {\bf 100}, no. 2, 026006 (2019).

  
  
  \bibitem{Narayan:2012ks} 
  K.~Narayan, T.~Takayanagi and S.~P.~Trivedi,
  ``AdS plane waves and entanglement entropy,''
  JHEP {\bf 1304}, 051 (2013)
  doi:10.1007/JHEP04(2013)051
  [arXiv:1212.4328 [hep-th]].
  
  
  
  
  \bibitem{Narayan:2013qga} 
  K.~Narayan,
  ``Non-conformal brane plane waves and entanglement entropy,''
  Phys.\ Lett.\ B {\bf 726}, 370 (2013)
  doi:10.1016/j.physletb.2013.07.061
  [arXiv:1304.6697 [hep-th]].
  

  \bibitem{Narayan:2015lka} 
  K.~Narayan,
  ``Lightlike limit of entanglement entropy,''
  Phys.\ Rev.\ D {\bf 91}, no. 8, 086010 (2015)
  doi:10.1103/PhysRevD.91.086010
  [arXiv:1408.7021 [hep-th]].
  
  
  \bibitem{Mukherjee:2014gia} 
  D.~Mukherjee and K.~Narayan,
  ``AdS plane waves, entanglement and mutual information,''
  Phys.\ Rev.\ D {\bf 90}, no. 2, 026003 (2014)
  doi:10.1103/PhysRevD.90.026003
  [arXiv:1405.3553 [hep-th]].
  
  
  
   \bibitem{Holzhey:1994we} 
  C.~Holzhey, F.~Larsen and F.~Wilczek,
  ``Geometric and renormalized entropy in conformal field theory,''
  Nucl.\ Phys.\ B {\bf 424}, 443 (1994).
  
  \bibitem{Calabrese:2004eu} 
  P.~Calabrese and J.~L.~Cardy,
  ``Entanglement entropy and quantum field theory,''
  J.\ Stat.\ Mech.\  {\bf 0406}, P06002 (2004).
  
  
  \bibitem{Casini:2009sr} 
  H.~Casini and M.~Huerta,
  ``Entanglement entropy in free quantum field theory,''
  J.\ Phys.\ A {\bf 42}, 504007 (2009)

  
  \bibitem{Ryu:2006bv} 
  S.~Ryu and T.~Takayanagi,
  ``Holographic derivation of entanglement entropy from AdS/CFT,''
  Phys.\ Rev.\ Lett.\  {\bf 96}, 181602 (2006).
  
    \bibitem{Ryu:2006ef} 
  S.~Ryu and T.~Takayanagi,
  ``Aspects of Holographic Entanglement Entropy,''
  JHEP {\bf 0608}, 045 (2006).
  
  \bibitem{Nishioka:2009un} 
  T.~Nishioka, S.~Ryu and T.~Takayanagi,
  ``Holographic Entanglement Entropy: An Overview,''
  J.\ Phys.\ A {\bf 42}, 504008 (2009).
  
  
  
  \bibitem{Bombelli:1986rw} 
  L.~Bombelli, R.~K.~Koul, J.~Lee and R.~D.~Sorkin,
  ``A Quantum Source of Entropy for Black Holes,''
  Phys.\ Rev.\ D {\bf 34}, 373 (1986).
  doi:10.1103/PhysRevD.34.373


\bibitem{Srednicki:1993im} 
  M.~Srednicki,
  ``Entropy and area,''
  Phys.\ Rev.\ Lett.\  {\bf 71}, 666 (1993)
  doi:10.1103/PhysRevLett.71.666
  [hep-th/9303048].


\bibitem{Casini:2003ix} 
  H.~Casini,
  ``Geometric entropy, area, and strong subadditivity,''
  Class.\ Quant.\ Grav.\  {\bf 21}, 2351 (2004)
  doi:10.1088/0264-9381/21/9/011
  [hep-th/0312238].
 
  
  \bibitem{Das:2005ah} 
  S.~Das and S.~Shankaranarayanan,
  ``How robust is the entanglement entropy: Area relation?,''
  Phys.\ Rev.\ D {\bf 73}, 121701 (2006)
  doi:10.1103/PhysRevD.73.121701
  [gr-qc/0511066].

  

   \bibitem{Myers:2012ed} 
  R.~C.~Myers and A.~Singh,
  ``Comments on Holographic Entanglement Entropy and RG Flows,''
  JHEP {\bf 1204}, 122 (2012)
  doi:10.1007/JHEP04(2012)122
  [arXiv:1202.2068 [hep-th]].
  
  
  \bibitem{SonStarinets} 
  D.~T.~Son and A.~O.~Starinets, ``Minkowski space correlators in AdS / CFT correspondence: Recipe and applications,'' JHEP {\bf 0209}, 042 (2002).
  
\bibitem{SonHerzog} 
  C.~P.~Herzog and D.~T.~Son, ``Schwinger-Keldysh propagators from AdS/CFT correspondence,'' JHEP {\bf 0303}, 046 (2003).
  
  \bibitem{BalasubramanianRoss} 
  V.~Balasubramanian and S.~F.~Ross, ``Holographic particle detection,'' Phys.\ Rev.\ D {\bf 61}, 044007 (2000).
  
\bibitem{Banks} 
  T.~Banks, M.~R.~Douglas, G.~T.~Horowitz and E.~J.~Martinec, ``AdS dynamics from conformal field theory,'' hep-th/9808016.  

\bibitem{Saddlepnt1} 
  J.~Louko, D.~Marolf and S.~F.~Ross, ``On geodesic propagators and black hole holography,'' Phys.\ Rev.\ D {\bf 62}, 044041 (2000).
  
  \bibitem{Balasubramanian:2011ur} 
  V.~Balasubramanian {\it et al.},
  Phys.\ Rev.\ D {\bf 84}, 026010 (2011)
  doi:10.1103/PhysRevD.84.026010
  [arXiv:1103.2683 [hep-th]].

    
\bibitem{Pedraza:2014moa} 
  J.~F.~Pedraza,
  ``Evolution of nonlocal observables in an expanding boost-invariant plasma,''
  Phys.\ Rev.\ D {\bf 90}, no. 4, 046010 (2014)
  doi:10.1103/PhysRevD.90.046010
  [arXiv:1405.1724 [hep-th]].
  
\bibitem{DiNunno:2017obv} 
  B.~S.~DiNunno, S.~Grozdanov, J.~F.~Pedraza and S.~Young,
  ``Holographic constraints on Bjorken hydrodynamics at finite coupling,''
  JHEP {\bf 1710}, 110 (2017)
  doi:10.1007/JHEP10(2017)110
  [arXiv:1707.08812 [hep-th]].

  
\bibitem{Kundu:2016dyk} 
  S.~Kundu and J.~F.~Pedraza,
  ``Aspects of Holographic Entanglement at Finite Temperature and Chemical Potential,''
  JHEP {\bf 1608}, 177 (2016)
  doi:10.1007/JHEP08(2016)177
  [arXiv:1602.07353 [hep-th]].

\bibitem{Chakrabortty:2020ptb} 
  S.~Chakrabortty, S.~Pant and K.~Sil,
  ``Effect of back reaction on entanglement and subregion volume complexity in strongly coupled plasma,''
  JHEP {\bf 2006}, 061 (2020)
  doi:10.1007/JHEP06(2020)061
  [arXiv:2004.06991 [hep-th]].


\end{thebibliography}
\end{document}